\documentclass{jpp}

\usepackage{epsfig}
\usepackage{graphicx}
\usepackage{epstopdf}

\usepackage{amsmath}
\usepackage{amssymb}
\usepackage{mathabx}

\usepackage{natbib}
\usepackage{color}
\usepackage{enumitem} 
\usepackage{tabu,multirow}

 \usepackage{soul}
\usepackage{color}

\newcommand\beq{\begin{equation}}
\newcommand\eeq{\end{equation}}

\newcommand{\ben}{\begin{eqnarray}}
\newcommand{\een}{\end{eqnarray}}
\newcommand{\benn}{\begin{eqnarray*}}
\newcommand{\eenn}{\end{eqnarray*}}

\newcommand{\apar}{ A_{\parallel}}

\newcommand{\pa}{\partial}

\newcommand{\lapp}{\Delta_\perp}

\newcommand{\gpar}{\nabla_{\parallel}}

\allowdisplaybreaks

\title{Direct kinetic Alfv\'en wave energy cascade in the presence of imbalance}

\author{T. Passot, P.L. Sulem and D. Laveder}

\affiliation{Universit\'e C\^ote d'Azur, Observatoire de la C\^ote d'Azur,  CNRS, Laboratoire J.L. Lagrange, Boulevard de l'Observatoire, CS  34229, 06304 Nice Cedex 4, France}

\begin{document}

\maketitle

\begin{abstract}
A two-field Hamiltonian gyrofluid model for kinetic Alfv\'en waves retaining ion finite Larmor radius corrections and parallel magnetic field fluctuations  is used to study direct turbulent cascades from the MHD to the sub-ion scales. For moderate energy imbalance and weak enough magnetic fluctuations, the spectrum of the transverse magnetic field and that of the most energetic wave display a steep transition zone near the ion scale, while the parallel transfer (and thus the parallel dissipation) remains weak. In this regime, the  perpendicular flux of generalized  cross helicity displays a significant decay past the ion scale, while the perpendicular energy flux remains almost constant. A phenomenological model suggests that  the interactions between co-propagative waves present at the sub-ion scales can play a central role in the development of a transition zone in the presence of a helicity barrier. 
\end{abstract}


\section{Introduction}

Alfv\'en wave (AW) turbulence, that plays an important role in magnetized collisionless plasmas such as the solar wind \citep{Bruno13, BC16}, is often imbalanced in the sense that the energy carried by the waves propagating in the forward and backward directions relatively to the ambient field have significantly different magnitudes \citep{Tu89,Lucek98,Wicks13}. The degree of imbalance is dependent on the type of wind \citep{Tu90} and also on the distance from the Sun \citep{Roberts87,Marsch-Tu90,Chen2020}. For example, in a classical model of solar wind, outgoing AWs are emitted at the Sun's surface, as a consequence of reconnection processes in the chromospheric magnetic network, while in-going waves result from reflection of these waves on density gradients \citep{Perez13,Chandran19}, 
or from parametric decay instability  \citep{Malara01}. In the case of low-frequency highly transverse perturbations, corresponding to the assumptions underlying the gyrofluid model derived in \citet{PST18}, imbalance is associated with a non-zero value of an invariant referred to as generalized cross helicity (GCH), which identifies with the cross-helicity at the magnetohydrodynamic (MHD) scales and with the magnetic helicity at sub-ion scales. As cross-helicity cascades to small scales and magnetic helicity to large scales \citep{Milo21}, the questions arises of the behaviour of the GCH flux arriving from large scales to the transition zone between MHD and sub-ion scales. Two publications recently addressed this issue. One uses a gyrofluid model (in the limit of small values of the electron beta parameter) which isolates the AW dynamics \citep{Meyrand21}. The other is based on a hybrid-kinetic PIC  simulation which enables the coupling of low-frequency quasi transverse AW with quasi-parallel high-frequency ion-cyclotron waves (ICW) \citep{Squire21}. The role of ICWs and of the dissipation associated with the proton cyclotron resonance at the spectral break separating the MHD and the sub-ion ranges was indeed stressed, using the data from the Messenger and Wind spacecraft missions \citep{Telloni15,Woodham18}.

In the regimes studied by \citet{Meyrand21}, the nonlinear parameter, defined as the ratio of the inverse nonlinear time to the Alfv\'en wave frequency is typically above unity. The dynamics then rapidly develops a significant transfer of energy towards the large parallel wavenumbers, resulting in the breaking of the asymptotics underlying gyrofluid models. Despite this shortcoming, the gyrofluid simulation reveals the existence of a helicity barrier which depletes the  energy flux towards the small perpendicular scales, leading to a transfer in the parallel direction at the ion scale, together with a steepening of the energy (and magnetic) spectra. Energy accumulates at  large scales and imbalance increases until saturation occurs under the action of a strong parallel dissipation. In contrast with usual fluid turbulence, the energy level that is reached depends on the magnitude of the parallel viscosity coefficient. In this framework, dissipation, which plays a central role in the dynamics, is not associated with any realistic physical phenomenon. The above scenario was then validated by the kinetic simulation of \citet{Squire21}, where the dissipation originates from ion-cyclotron resonance. In this simulation, performed at $\beta_e =0.3$, the level of fluctuations remains moderate because cyclotron resonance is rapidly triggered. In fact, an argument based on critical balance, together with the requirement that $k_\| d_i=1$ (where $d_i$ is the ion inertial length) at perpendicular scales such that $k_\perp \rho_i=1$ ($\rho_i$ denoting the ion Larmor radius), leads to a saturation amplitude scaling like $\beta_i^{1/2}$. As a consequence, when $\beta_e$  is taken of order unity (for comparable ion and electron temperatures), one may expect that high amplitude fluctuations should develop before a stationary regime can establish. Recent observations using Parker Solar Probe have comforted the results of the above simulation, with evidence of the coincidence between the presence of ICWs and of magnetic spectra displaying an ion transition zone associated with a polarity change, as expected when Alfv\'en waves trigger the formation of ICWs \citep{Bowen21}.

Transition zones are commonly observed in the solar wind at 1AU, where the plasma beta parameter is of order unity or larger (see e.g. \citet{Huang_2021} and references therein). When the amplitude of the fluctuations is moderate (such as in the fast wind), the above argument suggests that ICWs can hardly be excited at the perpendicular ion scale. In this context, an alternative mechanism should be at work, like for example the effect of co-propagating Alfv\'en wave interactions that become relevant at ion scales \citep{Voitenko16,Voitenko20}. In order to address this question, it is appropriate to use a two-field gyrofluid model \citep{PST18}, where driving and dissipation have been supplemented, which only retains the Alfv\'en waves, but requires that the fluctuation amplitudes remain small and the parallel dynamics sub-dominant (conditions that can be checked a posteriori).  We chose to prescribe the "energy imbalance" (referred to as imbalance in the following) by maintaining constant the amplitudes of the largest modes retained in the simulations, a driving process classically used in  hydrodynamics turbulence \citep{Siggia78}. It turns out that, in the quasi-stationary phases of the simulation, the GCH and energy injection rates are roughly constant  (equal to the total dissipation rates), the "injection imbalance", measured by their ratio, resulting from the dynamics.

The paper is organized as follows. Section \ref{Gyrofluid} describes the model and the numerical setup. Section \ref{GCH-barrier} discusses the gyrofluid simulations, showing that, for moderate imbalance, the spectrum of the magnetic fluctuations (and also that of the most energetic wave) displays a steep transition zone near the ion scale, while the parallel transfer (and thus the parallel dissipation) remains weak. The helicity barrier is mostly conspicuous on the perpendicular GCH flux, which displays a significant decay past the ion scale, while the transverse energy flux remains almost constant. In the situation where parallel energy and cross-helicity dissipations are negligible, the effect of co-propagating wave interactions seems to be the only alternative mechanism at the origin of the transition zone within the context of the present model. Section \ref{transition-zone} discusses this mechanism using a Leith-type  phenomenological spectral model. Section \ref{conclusion} is the conclusion.

\section{Equations and numerical setup} \label{Gyrofluid}
 Let us consider a situation where the  plasma dynamics results from weak perturbations of a homogeneous equilibrium state 
 characterized by a  density $n_0$, isotropic ion and electron temperatures $T_{0i}$ 
 and $T_{0e}$ (characterized by their ratio $\tau = T_{0i}/T_{0e}$), and subject to an ambient magnetic field of amplitude $B_0$
 along the $z$-direction. The various characteristic  scales are conveniently measured in terms of the  
 sonic Larmor radius $\rho_s  = c_s/\Omega_i$, where $c_s = \sqrt {T_{0e}/m_i}$ is the sound speed
 and $\Omega_i= eB_0/(mc)$ the ion gyrofrequency. 
 In particular, the ion Larmor
 radius  $\rho_i =v_{th\,i}/\Omega_i$ (where the ion thermal velocities is given by 
 $v_{th , i}=(2T_{0i}/m_i)^{1/2}$) and the  ion inertial length  $d_i=v_A/\Omega_i$ (where $v_A=B_0/(4\upi n_0 m_i)^{1/2}
 =c_s \sqrt{2/\beta_e}$ is the Alfv\'en velocity), satisfy
 $\rho_i = \sqrt{2\tau} \rho_s$ and 
 $d_i = \sqrt{2/{\beta_e}} \rho_s$.
 
The two-field gyrofluid model is  derived from the full gyrofluid model of \citet{Bri92} under specific scaling assumptions. Such a simple reduced fluid model does not retain physical phenomena like kinetic dissipation, but also misses some nonlinear couplings with other modes. Discarding these effects, one here focuses on the  nonlinear dynamics of AWs, or kinetic Alfvén waves (KAWs). Fully kinetic simulations will ultimately be necessary to validate the findings.  Like the parent model, the model is not restricted to the small $\beta_e$ regime, unless physics at the electron inertia scale $d_e$ is retained (in the absence of electron FLR terms in the two-field equations, the electron Larmor radius must indeed remain  smaller than $d_e$). More precisely, leaving aside issues related to kinetic damping, the model reproduces Reduced MHD in the limit of large transverse scales, and Electron Reduced MHD at sub-ion scales, both systems describing AWs or KAWs, uncoupled to slow waves, with no restriction on $\beta_e$  (especially since the parallel magnetic field component is retained). Only at the ion scales, where the system includes a dispersive term associated with the electron pressure gradient, can a coupling between Alfvén and slow waves become relevant (see Appendix E of \citet{Schekochihin09}). We here neglect the electron inertia, so the model covers a spectral range extending from the MHD to the sub-ion scales, within an anisotropic scaling where the transverse scales are much smaller than the parallel ones. 

 \subsection{The two-field gyrofluid model}
The model (which retains no dissipation process) is written as equations for the electron gyrocenter number density $N_e$ and the parallel component of the magnetic potential $A_\|$. When electron inertia is neglected, it takes the form  
\begin{eqnarray}
&&\partial_t N_e +[\varphi,N_e]-[B_z,N_e]+\frac{2}{\beta_e}\nabla_\| \Delta_\perp A_\|=0\label{eq:gyro-2fields-Ne} \label{eq:Ne}\\
  &&\partial_t A_\| 
 + \nabla_\| (\varphi-N_e-B_z)=0\label{eq:gyro-2fields-A}.\label{eq:A}
\end{eqnarray}
Here, $\Delta_\perp = \partial_{xx} + \partial_{yy}$ is the Laplacian in the plane transverse to the ambient field and  $[f,g]= \partial_x f \partial_y g-\partial_y f \partial_x g$  the canonical bracket of two scalar functions $f$ and $g$.
Furthermore, $\Gamma_n$ denotes the (non-local) operator 
$\Gamma_n(-\tau \Delta_\perp)$ which in Fourier space reduces to the multiplication by the function  $\Gamma_n(\tau k_\perp^2)$, defined by  $\Gamma_n(x) = I_n(x) e^{-x}$ where $I_n$ is the modified Bessel function of first type of order n. For a scalar function $f$, the parallel gradient operator $\gpar$ is defined by $\gpar f=-[\apar , f]+ \pa_z f$.

The equations are written in a nondimensional form, using the following units: $\Omega_i^{-1}$ for time, $\rho_s$ for the space coordinates (and thus $\rho_s^{-1}$ for the wavenumber components),  $B_0$ for the parallel magnetic fluctuations $B_z$, 
$n_0$ for the electron gyrocenter density $N_e$, $T_e/e$ for the electric potential $\varphi$ and $B_0\rho_s$ for the parallel magnetic potential $A_\|$.

Introducing the operators 
$L_1 = {2}/{\beta_e}  +(1+2\tau)(\Gamma_0-\Gamma_1)$,
$L_2 = 1 + (1-\Gamma_0)/\tau - \Gamma_0 +\Gamma_1$,
$L_3 = (1-\Gamma_0)/\tau$ and 
$L_4 = 1-\Gamma_0+\Gamma_1$, 
the parallel magnetic fluctuations are given by $ B_z = M_1 \varphi$, with $M_1 =  L_1^{-1} L_2 $, and the electrostatic potential by 
$\varphi =-M_2^{-1} N_e$, where $M_2 = L_3 + L_4L_1^{-1}L_2$ is positive definite. Thus, $B_z$ and $\varphi$  can both be expressed in terms of $N_e$.

At the linear level, the phase velocity $v_{ph}= \omega/k_z$ is given by 
\begin{equation}
v_{ph}^2\equiv\left(\frac{\omega}{k_z}\right)^2 = 
\frac{2}{\beta_e} k_\perp^2 \frac{1 -{\widehat M}_1 + {\widehat M}_2} {{\widehat M}_2},
\end{equation}
where the caret refers to the Fourier symbol of the operator.
The associated operator $V_{ph}$ is given by 
\begin{equation}
	V_{ph}= s (-\lapp)^{1/2} (1 -M_1 + M_2)^{1/2}M_2^{-1/2},
\end{equation}
where  $s= (2/\beta_e)^{1/2}$ is the equilibrium Alfv\'en velocity in sound speed units.
In physical space, the  eigenmodes, that can be viewed as generalized Elsasser potentials,  are given by $\mu^\pm=\Lambda \varphi \pm s A_\|$,
where $\Lambda = (-\lapp)^{-1} (1+M_2-M_1)^{1/2} M_2^{1/2}$.

In the absence of dissipation processes (needed in the turbulent regime), the system (\ref{eq:Ne})-(\ref{eq:A})  preserves the energy ${\mathcal E}$ and the generalized cross-helicity ${\mathcal C}$,
\begin{eqnarray}
&&{\mathcal E} = \frac{1}{2} \int \Big ( \frac{2}{\beta_e} |\bnabla_\perp A_\| |^2 
+ \frac{4\delta^2}{\beta_e^2}|\Delta_\perp A_\| |^2 
- N_e(\varphi -N_e-B_z) \Big ) d^3 {x}, \label{energy}\\
&&{\mathcal C} =-\int N_e  A_\|d^3 {x}.\label{defC}
\end{eqnarray}

In terms of $\mu^\pm$, the two invariants rewrite ($D=(-\Delta)^{1/2}$), 
\begin{eqnarray}
{\mathcal E}  &=&  \frac{1}{4} \int \left \{ (D \mu^+)^2 + (D \mu^-)^2 \right \} d^3x\\
{\mathcal C} &=& \frac{1}{4} \int V_{ph}^{-1}\{D \mu^+)^2 - (D \mu^-)^2 \} d^3x.
\end{eqnarray}

\subsection{Numerical setup}
Equations (\ref{eq:Ne})--(\ref{eq:A}) with $\beta_{\rm e}=2$ and $\tau = 1$  were solved in a isotropic periodic box of size $L$, using a spectral method for the space variables and a third-order Runge-Kutta scheme for the time stepping. An eight-order hyper-dissipation operator $\nu_\perp(\Delta_\perp)^4 + \nu_z \partial^8_{z}$ acting on $N_e$ or on $A_\|$ is supplemented in the equations for the corresponding quantities. For convenience, the parameters $\nu_\perp$ and $\nu_\|$ are referred to as transverse (or perpendicular) and parallel viscosities. The system is driven by freezing the amplitude of the modes whose transverse wavenumber stands in the first spectral shell and the parallel one corresponds to $k_z (L / 2\upi)= \pm 1$, as in \citet{Siggia78}. A main interest of this procedure is that it permits simulations with a high imbalance and a moderate nonlinearity parameter. In such a regime, the injection rates of energy and of GCH are not prescribed, in contrast with the situations where the system is driven by an external random force or by a negative damping  \citep{Meyrand21}.

\begin{table}
\begin{center}
\def~{\hphantom{0}}
\begin{tabular}{cccccccccc}

  &     &            &   &          &                    &              &                    &            &  \\
  & $N$ & $L/2 \upi$ & I & $\chi_0$ & $\chi_{\rm final}$ & $\nu_{perp}$ & $\nu_z/\nu_{perp}$ & $\Delta t$ & Color code  \\  
       
\hline
\hline
   
$R_1$ & \hspace{0.3cm} $784^3$ & \hspace{0.3cm} 11 & \hspace{0.3cm} 100 &  \hspace{0.3cm} 0.47 & \hspace{0.3cm} 0.61 & \hspace{0.3cm} $3\cdot 10^{-9}$ & \hspace{0.3cm} 1  & \hspace{0.3cm} $2\cdot 10^{-3}$    \\ 

\hline

$R_{2}$ & \hspace{0.3cm} $280^3$ & \hspace{0.3cm} 5.5 & \hspace{0.3cm} 100 & \hspace{0.3cm} 0.50 & \hspace{0.3cm} 0.68 & \hspace{0.3cm} $5\cdot 10^{-9}$ & \hspace{0.3cm} 1  & \hspace{0.3cm} $3.125\cdot 10^{-3}$ & red   \\

\hline

$R_{3}$ & \hspace{0.3cm} $280^3$ & \hspace{0.3cm} 5.5 & \hspace{0.3cm} 10 & \hspace{0.3cm} 0.43 & \hspace{0.3cm} 0.60 & \hspace{0.3cm} $2\cdot 10^{-8}$ & \hspace{0.3cm} 1  & \hspace{0.3cm} $3.125\cdot 10^{-3}$ & green   \\

\hline

$R_{4}$ & \hspace{0.3cm} $280^3$ & \hspace{0.3cm} 5.5 & \hspace{0.3cm} 1 & \hspace{0.3cm} 0.36 & \hspace{0.3cm} 0.56 & \hspace{0.3cm} $5\cdot 10^{-8}$ & \hspace{0.3cm} 1  & \hspace{0.3cm} $3.125\cdot 10^{-3}$ & orange   \\

\hline

$R_{5}$ & \hspace{0.3cm} $280^3$ & \hspace{0.3cm} 5.5 & \hspace{0.3cm} 100 & \hspace{0.3cm} 0.87 & \hspace{0.3cm} 1.2 & \hspace{0.3cm} $2\cdot 10^{-8}$ & \hspace{0.3cm} 1  & \hspace{0.3cm} $3.125\cdot 10^{-3}$ & magenta   \\

\hline

$R_{6}$ & \hspace{0.3cm} $280^3$ & \hspace{0.3cm} 5.5 & \hspace{0.3cm} 100 &  \hspace{0.3cm} 0.50 & \hspace{0.3cm} 0.45& \hspace{0.3cm} $5\cdot 10^{-9}$ & \hspace{0.3cm} 0.025  & \hspace{0.3cm} $3.125\cdot 10^{-3}$ & blue   \\

\hline

$R_{7}$ & \hspace{0.3cm} $280^3$ & \hspace{0.3cm} 5.5 & \hspace{0.3cm} 100 & \hspace{0.3cm} 0.50 & \hspace{0.3cm} 0.43 & \hspace{0.3cm} $5\cdot 10^{-9}$ & \hspace{0.3cm} 0  & \hspace{0.3cm} $3.125\cdot 10^{-3}$ & cyan   \\

\hline

$R_{8}$ & \hspace{0.3cm} $720^3$ & \hspace{0.3cm} 5.5 & \hspace{0.3cm} 1400 &  \hspace{0.3cm} 0.49 & \hspace{0.3cm} 0.55 & \hspace{0.3cm} $2\cdot 10^{-10}$ & \hspace{0.3cm} 1  & \hspace{0.3cm} $8\cdot 10^{-4}$& violet    \\

\hline 
\hline 
\end{tabular}
\label{tabruns}
\caption{Parameters of the runs, together with the values $\chi_0$ and $\chi_{\rm final}$ of the nonlinear parameter at the beginning and at the end of the simulation, estimated by the square root of the magnetic energy at this time. The color code refers to the color of the lines in figures \ref{fig7}-\ref{fig11}.}
\end{center}
\end{table}

\section{The GCH barrier}\label{GCH-barrier}

\subsection{Spectral transition zone and arrest of the GCH cascade}\label{GCH-arrest}

In addition to the imbalance given by the ratio $I={\mathcal E}^+ / {\mathcal E}^+$ of the energies of the forward and backward propagating waves,  another quantity  governing the dynamics is provided by the nonlinearity parameter $\chi$, defined as the ratio of the inverse nonlinear time to the Alfv\'en frequency. 
At  early times, when the maintained modes are dominant, a simple phenomenological argument \citep{Milo20} leads to  $\chi \approx (k_{\perp 0}/k_{z 0}) |B_\perp(k_{\perp 0})| = |B_\perp(k_{\perp 0})|$,  where $B_\perp$ is the transverse magnetic fluctuations (measured in units of the ambient field), and the index zero refers to the maintained modes for which $k_{\perp 0} = k_{z 0}$ in the (anisotropic) units of the gyrofluid model. At  later times, in  the  isotropic simulation box, the rms value of the magnetic fluctuation is used as an estimation of $\chi$.
In this paper, we concentrate on the case where the nonlinearity parameter keeps moderate values (initially $\chi \leq 0.5$), unless otherwise stated.

\begin{figure}
\centerline{
\includegraphics[width=0.5\textwidth]{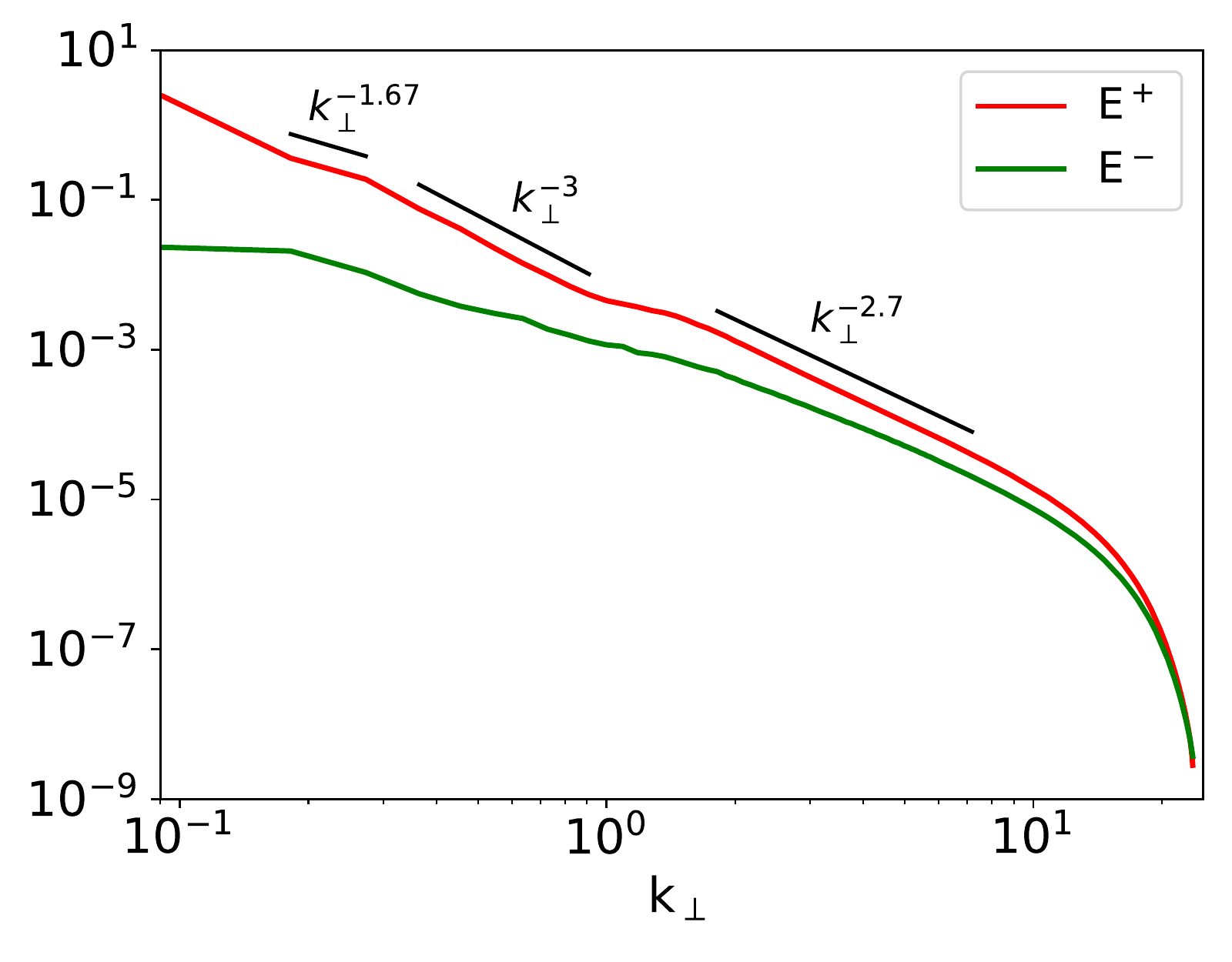}
\includegraphics[width=0.5\textwidth]{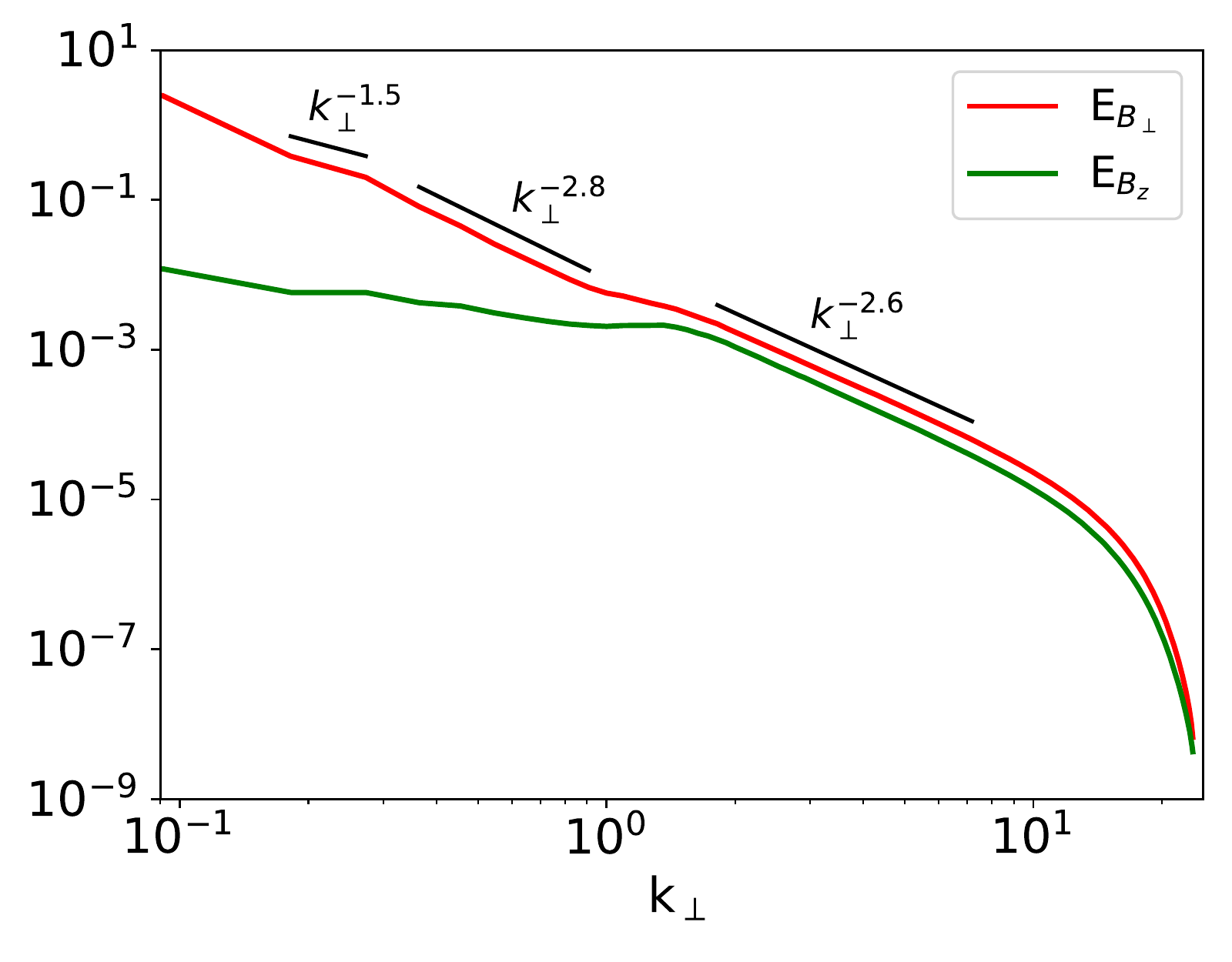}
}
\caption{Transverse energy spectrum $E^\pm$ of the forward and backward propagating waves (left panel) and energy spectrum of the transverse and parallel magnetic fluctuations $E_{B_\perp}$ and $E_{B_z}$, averaged on three Alfv\'en crossing times $L/v_A$, for run $R_1$.}
\label{fig1}
\end{figure}

\begin{figure}
\centerline{
\includegraphics[width=0.5\textwidth]{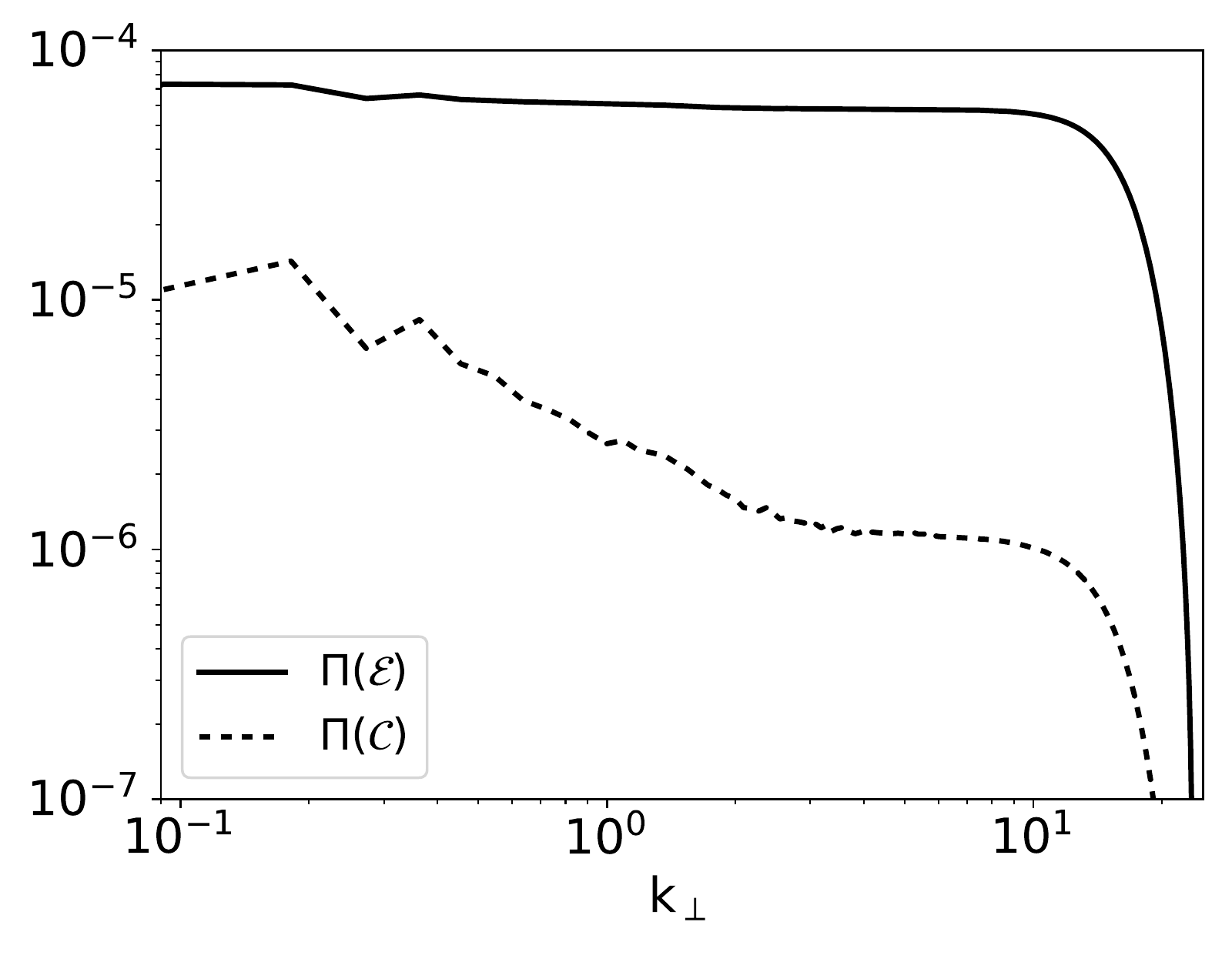}
}
\caption{Energy and GCH fluxes in the transverse direction for run $R_1$ averaged in the same time interval as in Fig. 1.}
\label{fig2}
\end{figure}

\begin{figure}
\centerline{
\includegraphics[width=0.5\textwidth]{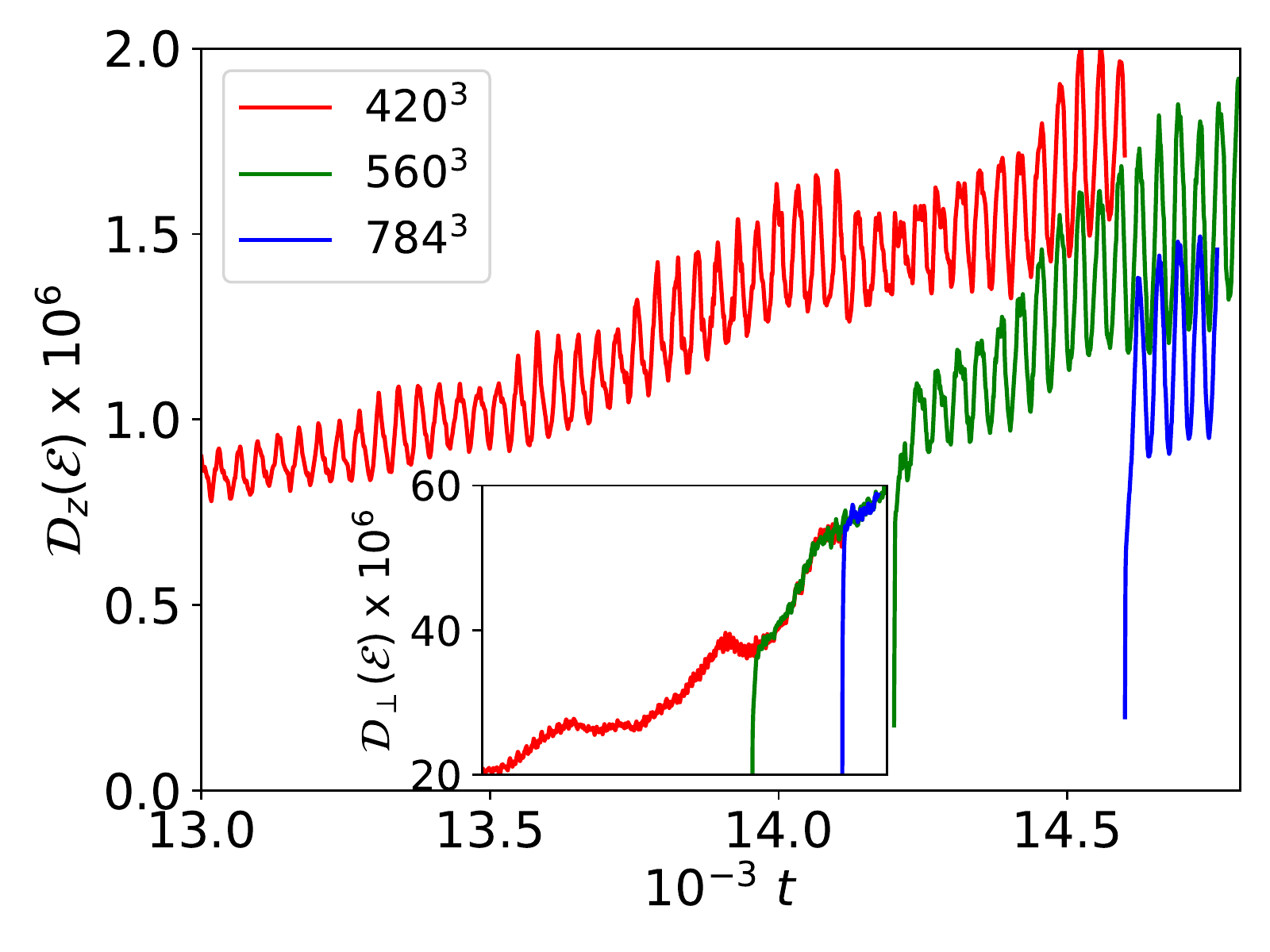}
\includegraphics[width=0.5\textwidth]{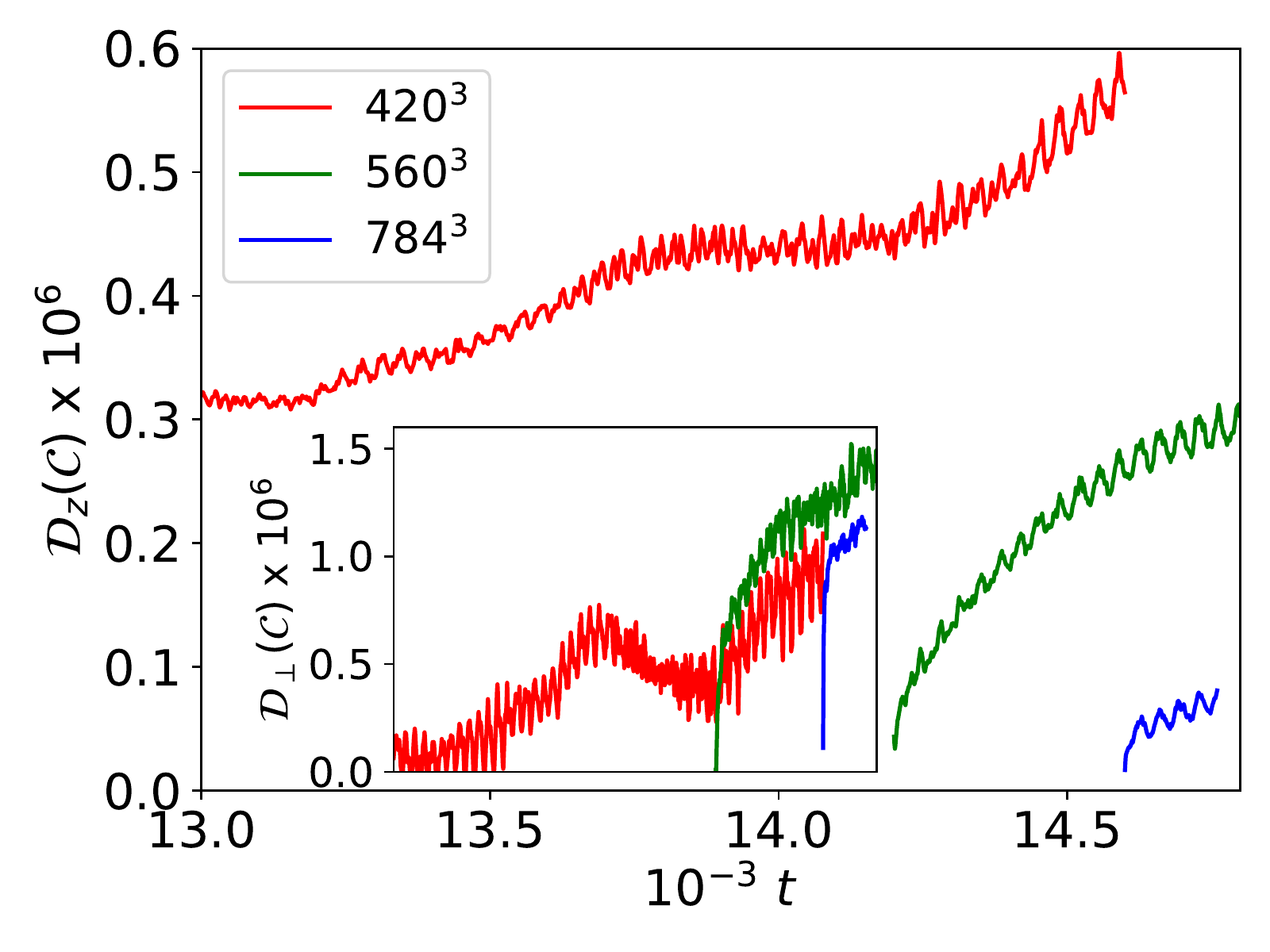}
}
\caption{Parallel energy  (left panel) and GCH (right panel) dissipations ${\mathcal D}_z({\mathcal E})$ and ${\mathcal D}_z({\mathcal C})$ respectively, for run $R_1$. The run was started using equal perpendicular and parallel viscosity coefficients $\nu=5 \times 10^{-8}$ using a resolution of $420^3$ collocation points (red line); a first restart was then performed with $\nu = 5 \times 10^{-9}$ and $560^3$ collocation points (green line), and another one with  $\nu = 3 \times 10^{-10}$ and $784^3$ collocation points (blue line). The inserts show the corresponding perpendicular energy and GCH dissipations ${\mathcal D}_\perp({\mathcal E})$ and ${\mathcal D}_\perp({\mathcal C})$.  Due to the lack of space, the time labels were not specified for the inserts that cover the same time interval as the main graphs.}
\label{fig3}
\end{figure}

We shall first illustrate our  main results on a large-resolution simulation (run $R_1$) that displays extended inertial ranges. At the end of this simulation, the system did not yet reach a  stationary state. Nevertheless, as mentioned in section \ref{sec:temporal_evolution}, the observed dynamics is similar to the one found in a lower-resolution simulation (run $R_2$), carried out until saturation is reached.
For the $R_1$ simulation characterized by an initial imbalance $I = 100$ and an initial value $\chi_0 = 0.47$ of the nonlinearity parameter, performed in a computational box of size $L=11 \times 2 \upi$, figure \ref{fig1} displays the transverse spectra (integrated over $k_z$) $E^\pm(k_\perp)$  of the forward and backward propagating waves (left panel), and   $E_{B_\perp}(k_\perp)$, $E_{B_z}(k_\perp)$ of the transverse and parallel magnetic fluctuations (right panel), averaged over the time interval [14656, 14760] corresponding to $1.5$  Alfv\'en crossing times  (see figure \ref{fig3} for a characterization of this interval relatively to the full simulation). 
The spectra  $E^+(k_\perp)$ and $E_{B_\perp}(k_\perp)$ display a transition zone at scales slightly larger than the sonic Larmor radius (corresponding to $k_\perp = 1$),  with a spectral index close to $-3$, located between a  sub-ion range with a spectral index $-2.7$ and a MHD range whose extension is  strongly limited by the numerical constraints which make difficult the simulation in  a box that would ideally cover a more extended range at large scale, while also retaining the small-scale dynamics. The transition zone is seen to extend up to the wavenumber where the spectrum $E_{B_z}(k_\perp)$ changes from a quasi-flat range to a decay range displaying the same spectral exponent  as $E_{B_\perp}(k_\perp)$. The large-scale $B_z$ fluctuations are much smaller than those of $B_\perp$ as expected when only Alfvén waves are excited at large scale (see e.g. \citet{kobayashi17}).

Figure \ref{fig2} shows the perpendicular fluxes of energy and GCH, averaged  over the same time interval as the spectra displayed in figure \ref{fig1}.
One observes that the energy flux is quasi-constant, even if the temporal fluctuations are comparable to the mean value (not shown) at scales larger than the transition one. Differently, the GCH flux undergoes a conspicuous decay at the transition range, a consequence of the  helicity barrier whose existence was pointed out by \citet{Meyrand21} in similar simulations (although in the asymptotic  regime $\beta_e \to 0$), performed  in the large $\chi$ regime where the energy flux to small transverse scales is also significantly inhibited by the barrier.

Because of numerical constraints, run $R_1$ was performed in three steps, associated with increased resolutions ($N=420^3$, $560^3$ and $784^3$ mesh points) and decreased (equal) parallel and perpendicular viscosities. This procedure permitted us to analyze the sensitivity of the parallel and perpendicular dissipations of energy (figure \ref{fig3} left) and GCH (figure \ref{fig3} right). We observe that the perpendicular dissipations are essentially not affected by the transition from one step to the following one, as expected in a usual turbulence cascade: reduction of the viscosity does not affect the dissipation and only shifts the dissipation range to smaller scales. In contrast, we observe that this is not the case for the parallel dissipation of energy and even more of GCH, which both decrease with the viscosity, confirming the absence of a standard turbulent cascade in the direction of the ambient field, at least in the regime of small or moderate nonlinear parameters considered in the present study.

\begin{figure}
\centerline{
\includegraphics[width=1.05\textwidth]{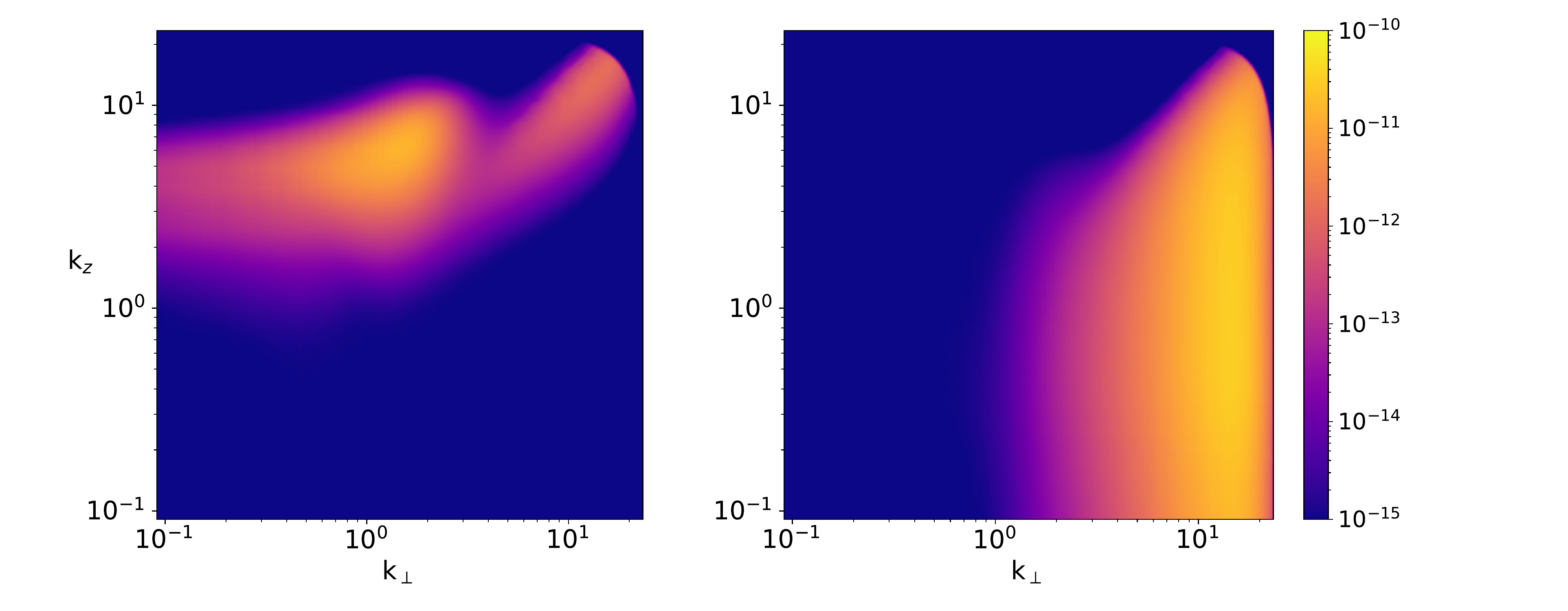}
}
\caption{Parallel (left) and perpendicular (right) dissipation of GCH in the $(k_\perp,k_z)$-plane (logarithmic coordinates) for run $R_1$.
}
\label{fig4}
\end{figure}

\begin{figure}
\centerline{
\includegraphics[width=1.05\textwidth]{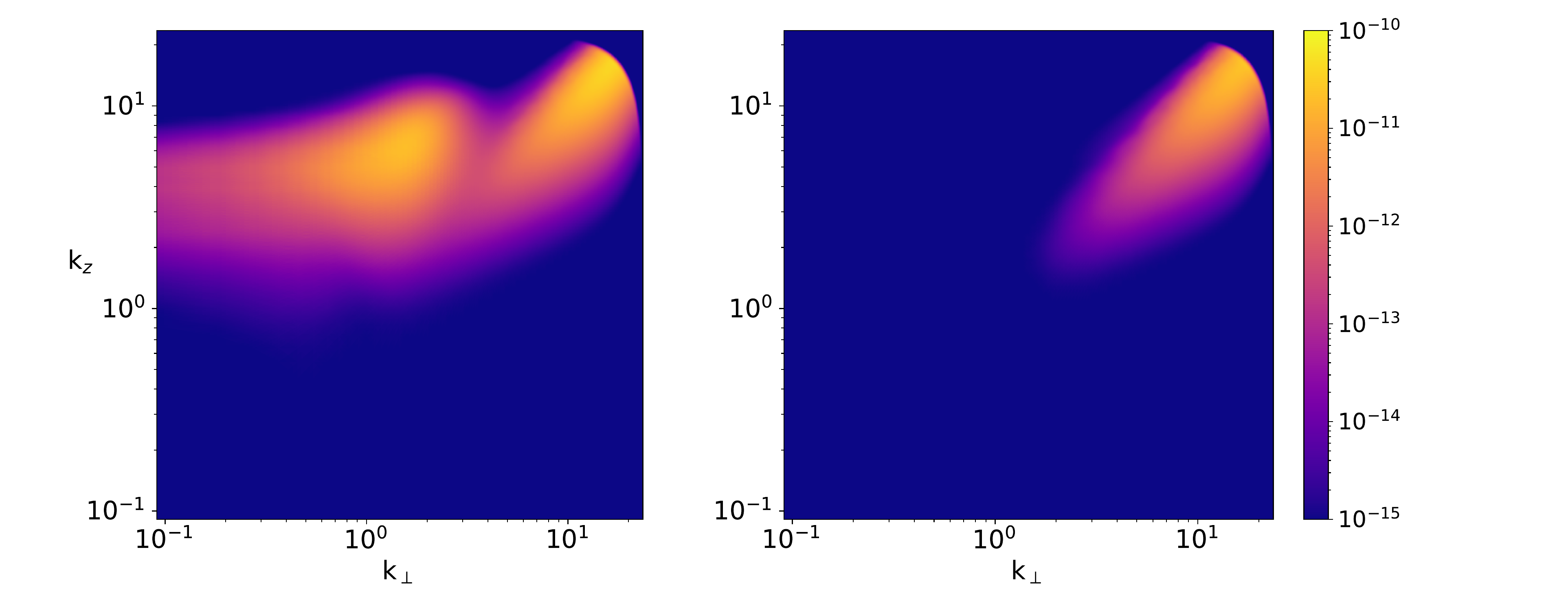}
}
\caption{Parallel dissipation of the energies of the $+$  (left) and $-$ (right) waves in the $(k_\perp,k_z)$-plane for run $R_1$.
}
\label{fig5}
\end{figure}

\begin{figure}
\centerline{
\includegraphics[width=0.5\textwidth]{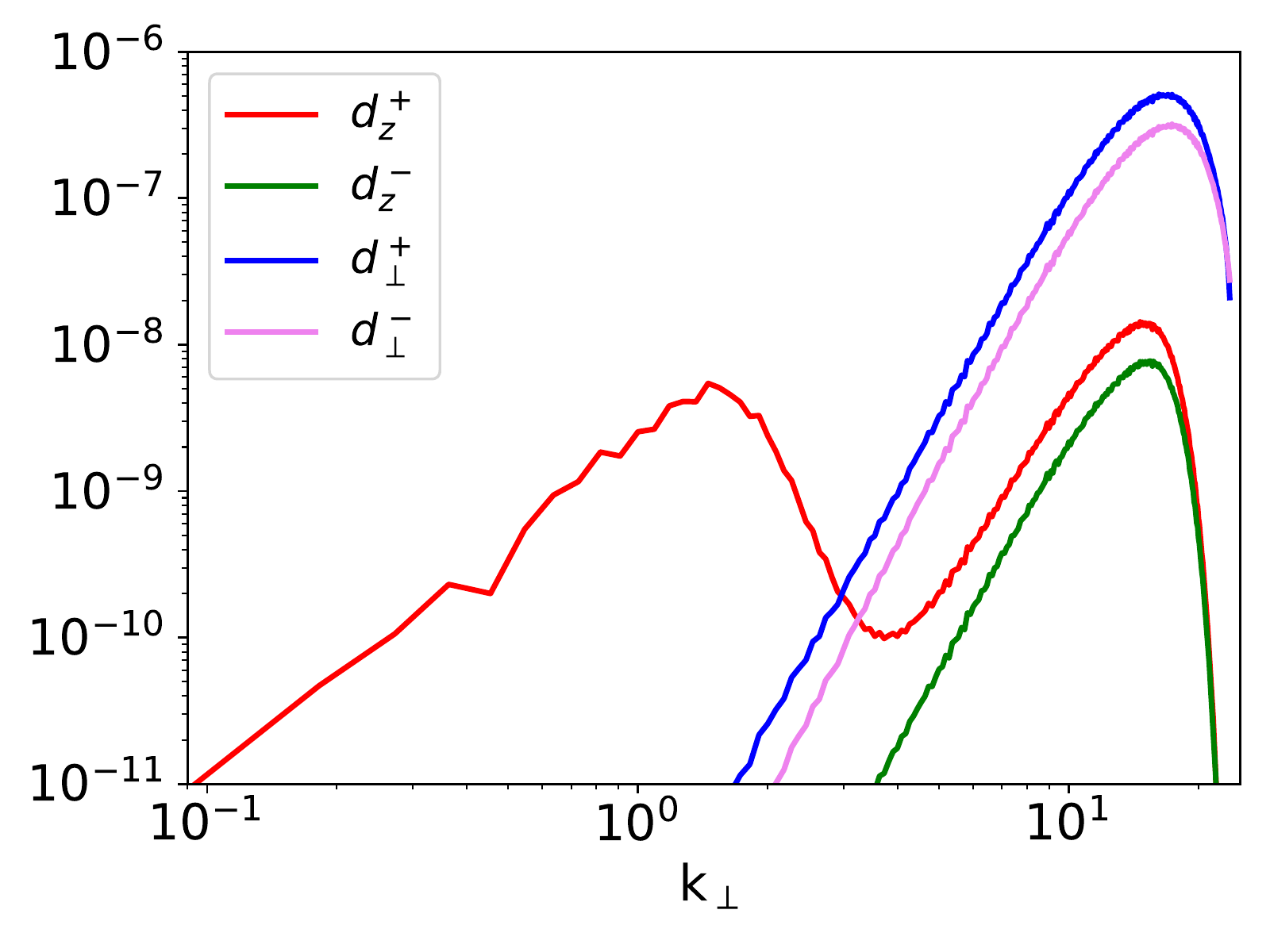}
\includegraphics[width=0.5\textwidth]{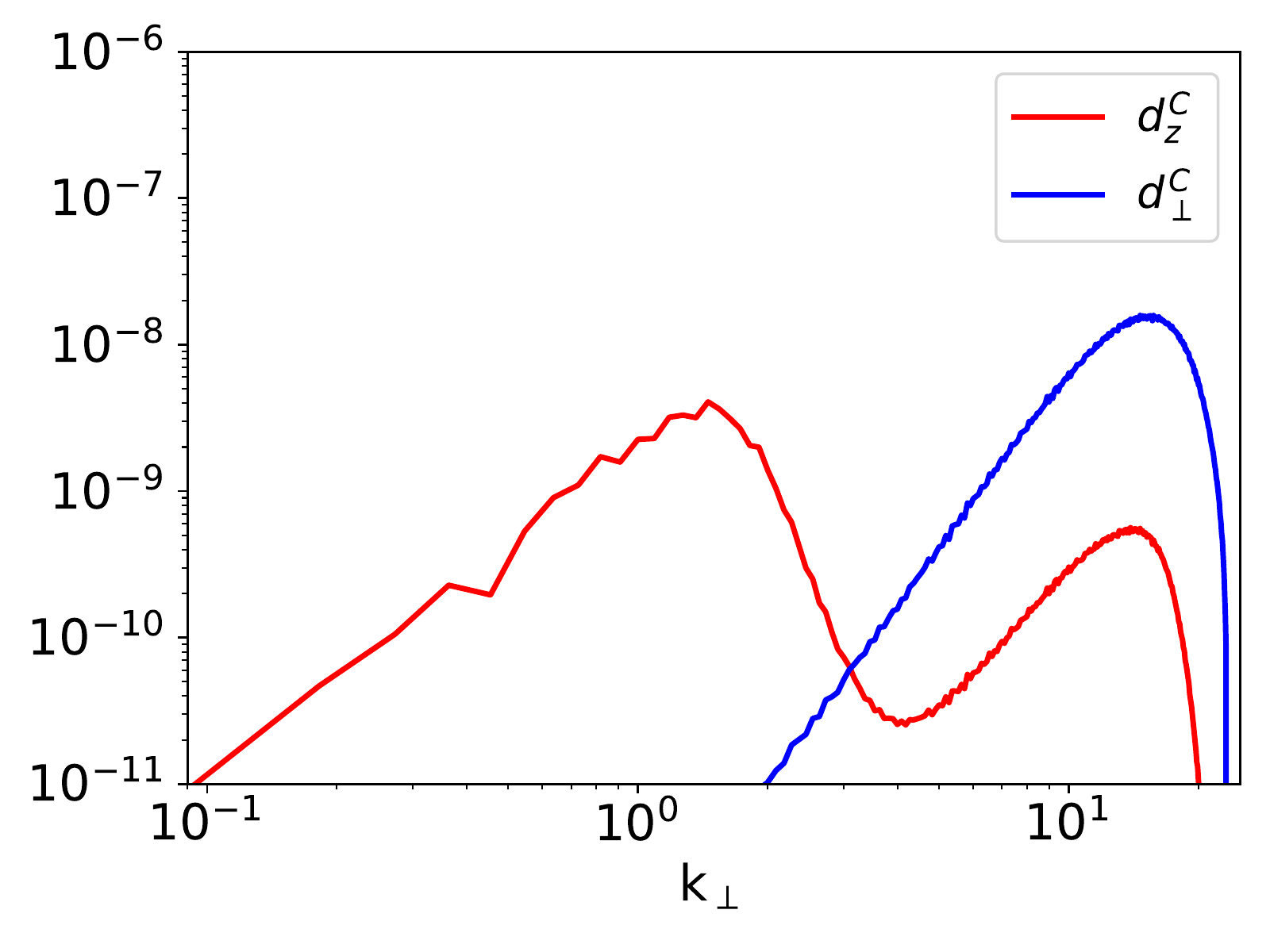}
}
\caption{Left panel: parallel (red) and perpendicular  (blue) dissipations of the energy of the $+$ wave, and parallel (green) and perpendicular (magenta) dissipations of the energy of the $-$ wave, integrated on $k_z$, versus $k_\perp$ for run $R_1$. Right panel: parallel (red) and perpendicular (blue) dissipations of GCH, integrated on $k_z$, versus $k_\perp$, for the same run.}
\label{fig6}
\end{figure}

Figure \ref{fig4} displays  two-dimensional color plots of the time-averaged parallel (left panel) and perpendicular (right panel) GCH dissipations, defined by $D_z^C(k_\perp, k_z) = \nu_z k_z^8 E_C(k_\perp, k_z)$ and $D_{\perp}^C(k_\perp, k_z) = \nu_\perp k_\perp^8 E_C(k_\perp, k_z)$ respectively, where both the viscosities $\nu_\perp$ and $\nu_z$ are here equal to $3 \times 10^{-10}$ (and where $E_C(k_\perp, k_z)$ stands for the  GCH power spectrum, after angle averaging in the transverse plane). In addition to the usual small-scale dissipation in both parallel and perpendicular directions,  one clearly observes the existence of a significant parallel dissipation of GCH, which is larger near a transverse wavenumber close to $k_\perp=1$  than at the smallest scales.

These observations can be understood by considering  the contributions of the $+$ and $-$ waves, separately.
Figure \ref{fig5} shows the two-dimensional  color  plots of the time-averaged parallel dissipations $D_z^+(k_\perp, k_z) = \nu_z k_z^8 E^+(k_\perp, k_z)$ and $D_z^-(k_\perp, k_z) = \nu_z k_z^8 E^-(k_\perp, k_z)$ of the energies associated with the forward and backward propagating waves. A dissipation spot is visible near $k_\perp = 1$  in the case of the  $+$waves only. In contrast, comparable dissipations of the forward and backward propagating waves  take place at  small scales, both in the parallel and perpendicular  directions. 
Indeed, the (weak) perpendicular transfer $\epsilon^{-}$ of $E^{-}$ is almost not affected by the GCH barrier, while $\epsilon^{+}$ is reduced past the barrier to a value close to $\epsilon^{-}$ since at these scales imbalance has almost disappeared (see also \citet{Meyrand21}, in a case where the helicity barrier affects the energy transfer more significantly than in the present simulation). In order to reach a stationary state, the transfer in the $z$-direction and the resulting parallel dissipation thus have to become  sizeable near the ion scale for the $+$ waves. At small scale, the dissipation of the two types of waves are comparable and  thus almost cancel out in the GCH dissipation, making the parallel dissipation of GCH to be dominant at transverse wavenumbers close to $k_\perp =1$, as observed in figure \ref{fig4}. 

A synthetic view of the dissipations in terms of the transverse wavenumber is obtained when integrating on $k_z$  the dissipations $D_z^+$, $D_z^-$, $D_{\perp}^+$, $D_{\perp}^-$ associated with the energy of the backward and forward propagating waves, and the analogous dissipations $D_z^C$, $D_{\perp}^C$ for the GCH. The resulting functions $d_z^+(k_{\perp})$, $d_z^-(k_{\perp})$, $d_{\perp}^+(k_{\perp})$, $d_{\perp}^-(k_{\perp})$, $d_z^C(k_{\perp})$, $d_{\perp}^C(k_{\perp})$ are displayed in figure \ref{fig6} (left panel: dissipation of energies; right panel: dissipations of GCH), under the same time averaging  as in figures \ref{fig4} and \ref{fig5}.
The perpendicular energy dissipation  is concentrated at small scales, with comparable magnitudes for both directions of propagation. Differently, for the parallel dissipation, while the energy of the backward propagating waves is dissipated at small scales only, the dissipation of the  forward propagating waves, and thus the GCH,  also displays a maximum near the transition zone. 

\begin{figure}
\centerline{
\includegraphics[width=0.5\textwidth]{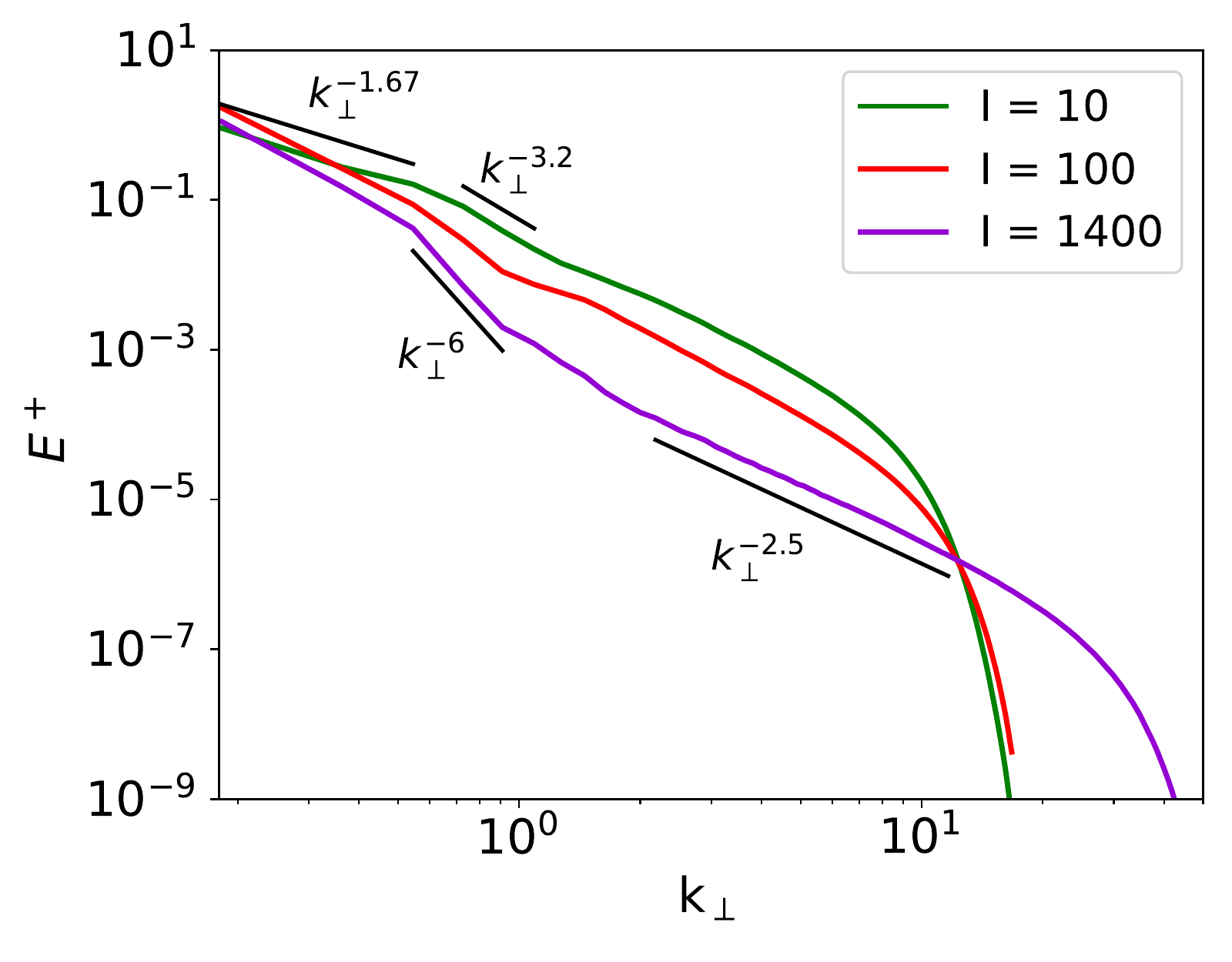}
\includegraphics[width=0.5\textwidth]{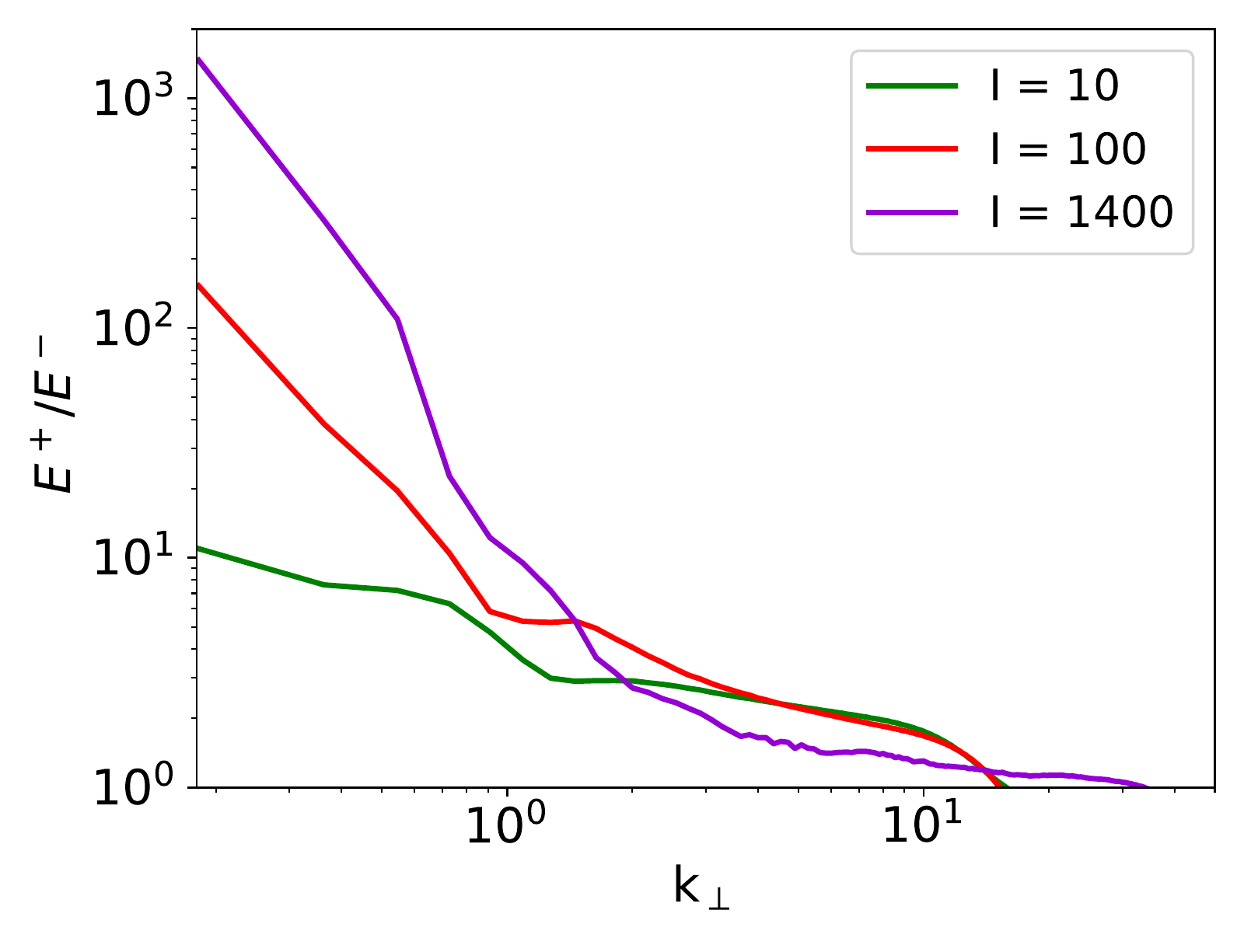}
}
\caption{Influence of the imbalance on the perpendicular $E^+(k_\perp)$ spectrum (left panel) and on  the ratio $E^+(k_\perp)/ E^-(k_\perp)$  (right panel): $I = 10$ (run $R_3$), $I=100$ (run $R_2$), $I=1400$ (run $R_8$).}
\label{fig7}
\end{figure}

It turns out that the effect of the helicity barrier is enhanced when the large-scale imbalance is increased, resulting in a steeper transition zone, in agreement with solar wind observations reported in \citet{Huang_2021}.
This is exemplified in figure \ref{fig7} (left) which displays the time-averaged spectrum $E^+(k_\perp)$ for three runs with different initial imbalance : $I=10$ (run $R_3$), $I=100$ (run $R_2$), $I=1400$ (run $R_8$). Time averages are performed in the intervals [20000, 25000], [30000, 40000] and [7064, 7096], respectively. Compared to run $R_1$, they have a similar value of $\chi$ but are integrated in a box of size $L=5.5\times 2\upi$, with resolutions of $280^3$ (for $I$ = 10, 100) and $720^3$ (for $I$ = 1400) mesh points. When considering the ratio $E^+(k_\perp)/E^-(k_\perp)$ (figure \ref{fig7}, right), we observe that it becomes close to unity for $k_\perp \gtrsim  2$, whatever the value of the imbalance at large scales.

\subsection{Temporal evolution of the dissipations and of the ideal invariants }
\label{sec:temporal_evolution}

\begin{figure}
\centerline{
\includegraphics[width=0.5\textwidth]{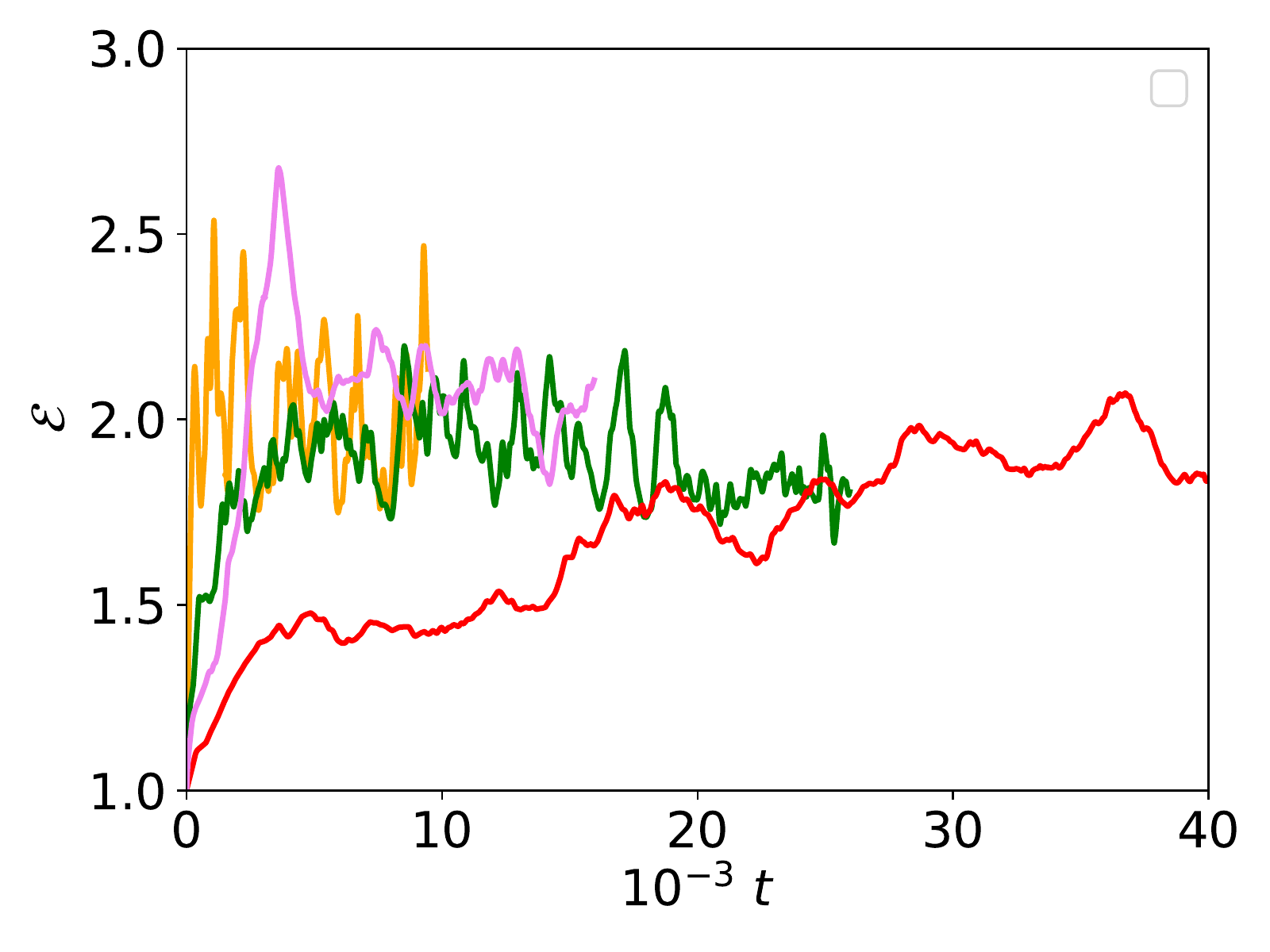}
\includegraphics[width=0.5\textwidth]{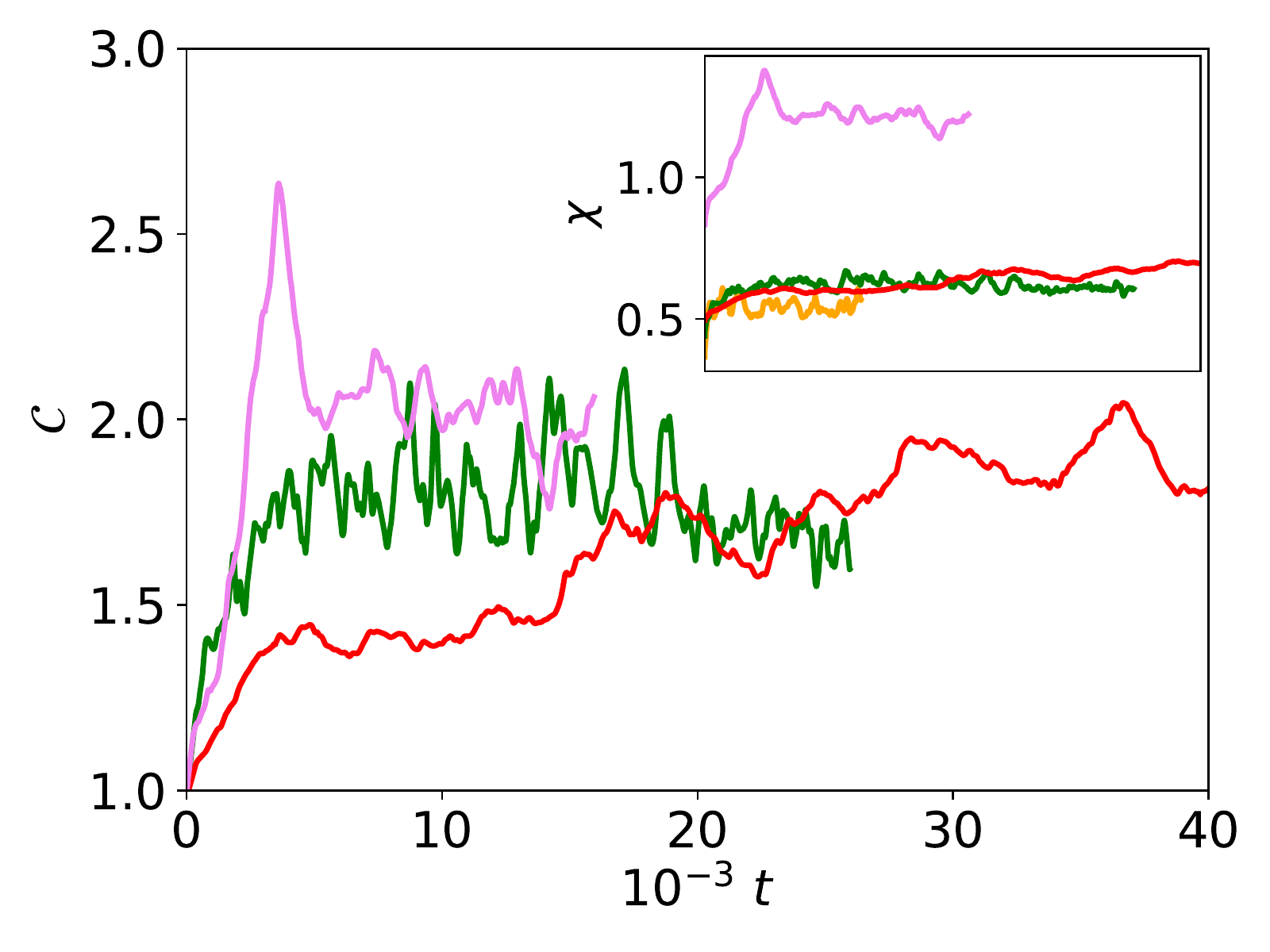}
}
\caption{Time evolution (averaged over 3 Alfvén crossing times) of the  energy (left panel) and GCH (right panel), normalized to their initial values, for  runs $R_2$, $R_3$ and $R_4$, with similar values of the nonlinear parameter $\chi$ (whose time  evolution is displayed in the insert within the right panel) and different imbalances: I=100 (red), I=10 (green), I=1 (orange), respectively; a fourth run ($R_5$) with I=100 and a larger  $\chi$ is also displayed in magenta.}
\label{fig8}
\end{figure}

While at zero imbalance the energy quickly reaches its saturation value (with nevertheless significant temporal oscillations), when the imbalance is increased, the time needed to reach saturation can become very large. 
This is shown in figure \ref{fig8} which also indicates that, for a small initial value of $\chi$, the typical saturation time increases with $I$ (runs $R_2$, $R_3$. $R_4$).\footnote{In order to perform simulations significantly longer than run $R_1$, we resorted to reduce the numerical resolution to $280^3$ mesh points and to use a smaller computational box, of size $L=5.5\times 2\upi$.}

\begin{figure}
\centerline{
\includegraphics[width=0.5\textwidth]{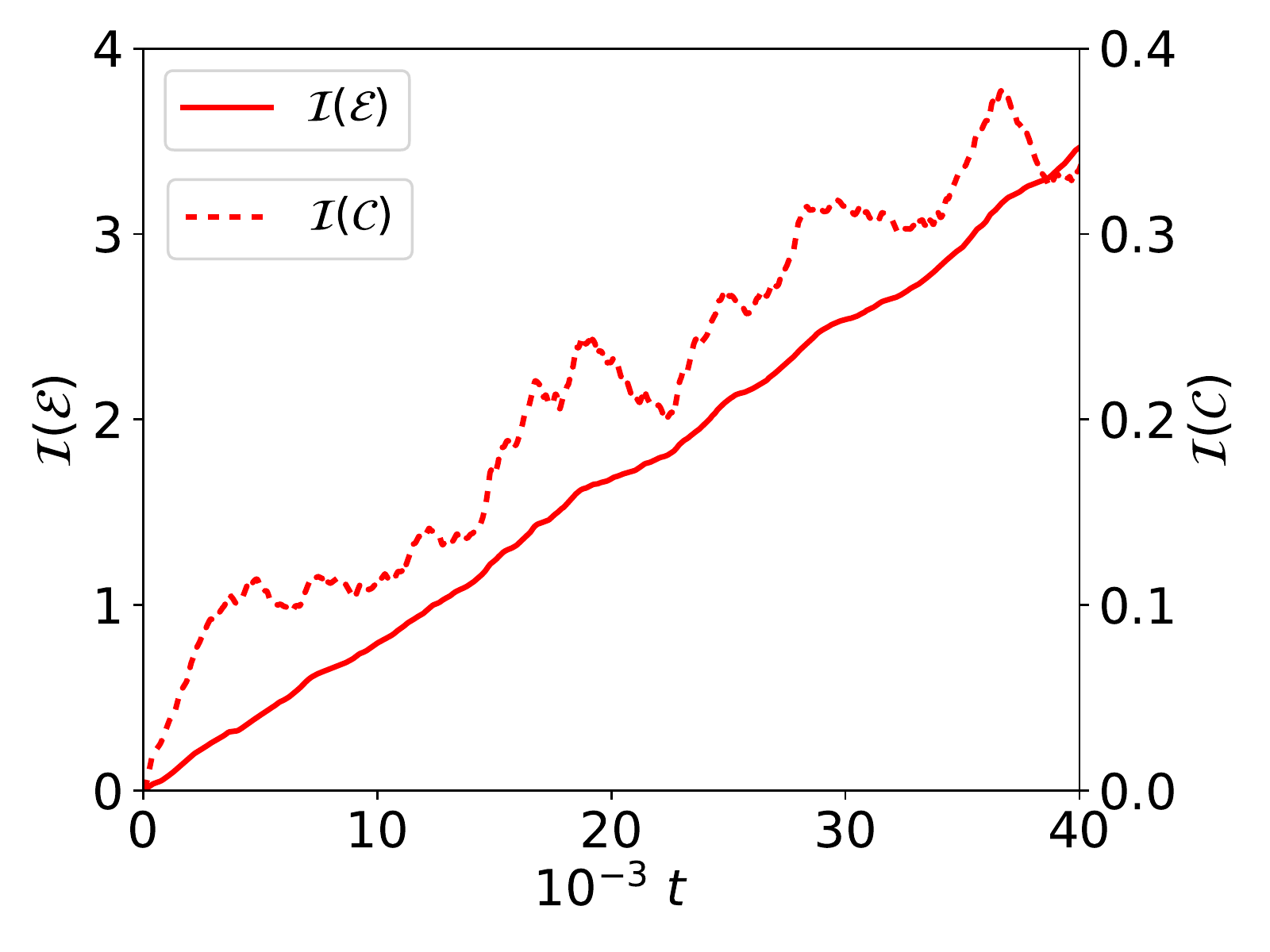}
\includegraphics[width=0.5\textwidth]{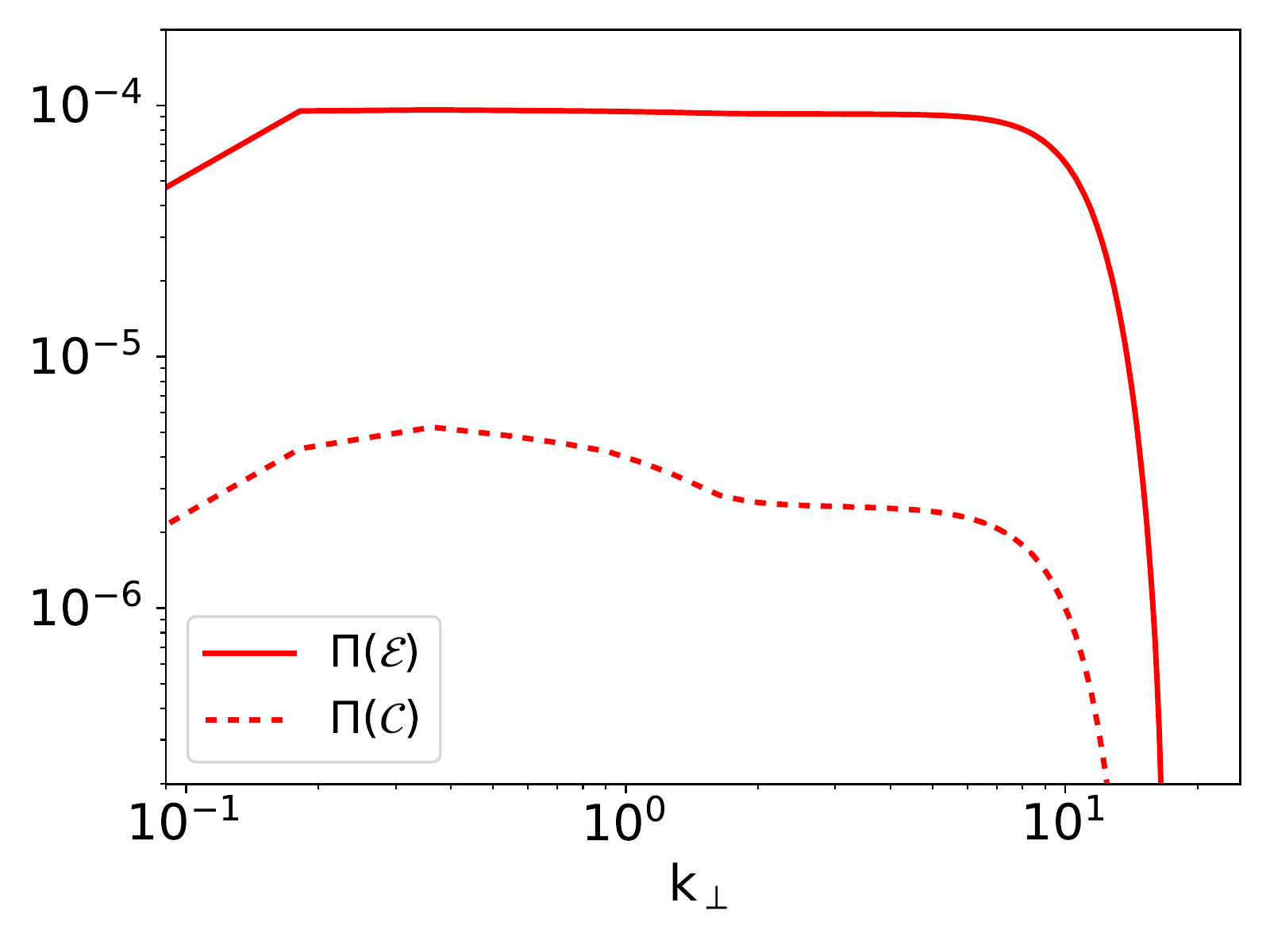}
}
\caption{Left: Cumulative injections of energy  (solid line) and  GCH (dashed line). Right: energy (solid line) and GCH (dashed line) fluxes of Run 2, averaged on the time interval $[30000, 40000]$.}
\label{fig9}
\end{figure}

Interestingly, for run $R_2$ with an energy imbalance $I=100$, the injection imbalance, given by the GCH to energy injection ratio, is  relatively small, with a value close to $0.1$ at saturation, as seen in figure \ref{fig9} (left). Here, the injection rates are given by the slope of the cumulative injections, estimated as the  sums of  the energy or the GCH and of their respective cumulative dissipations. Figure \ref{fig9} (right) displays the energy and GCH fluxes for run $R_2$ in the  steady state.
They look qualitatively similar to those of  run $R_1$, and in particular confirm the persistence of the GCH barrier in the saturated regime, in spite of the small value of the injection imbalance. In contrast, an energy barrier is hardly visible in this regime. 

\begin{figure}
\centerline{
\includegraphics[width=0.5\textwidth]{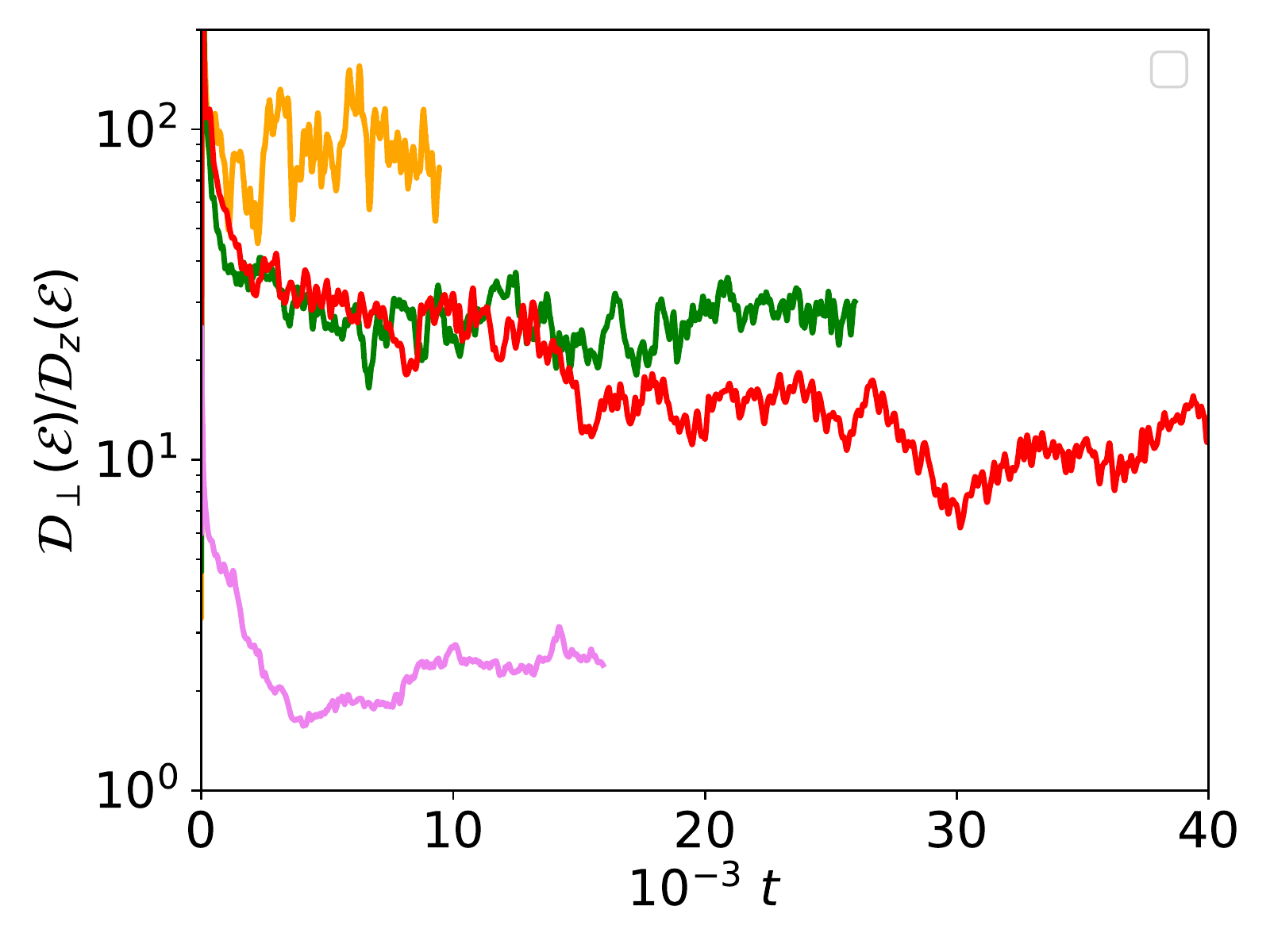}
\includegraphics[width=0.5\textwidth]{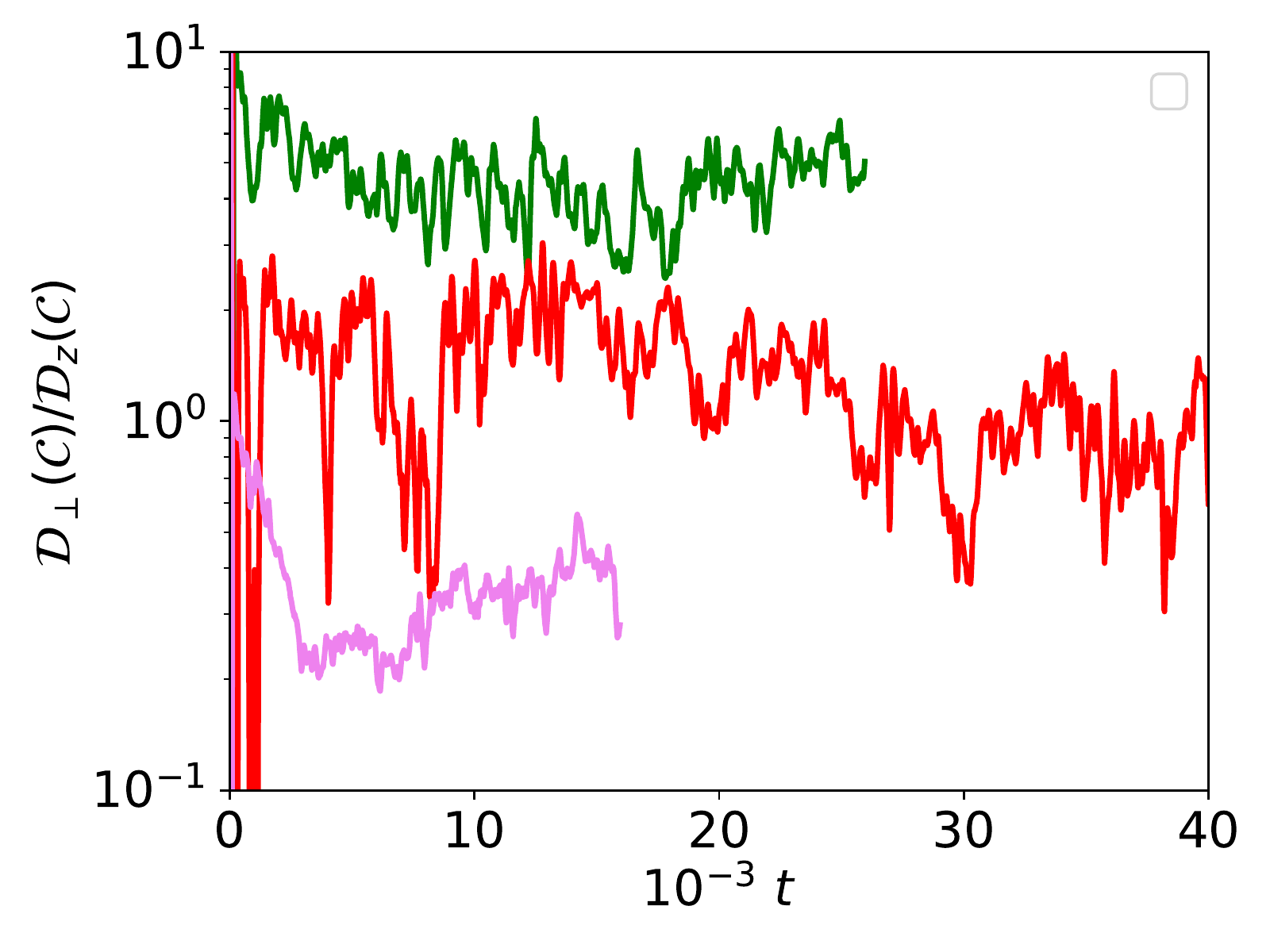}
}
\caption{Time evolution (averaged over 3 Alfvén crossing times) of the ratio between perpendicular and parallel dissipations of  energy (left) and GCH (right), for the  simulations shown in figure \ref{fig8} (same color code). }
\label{fig10}
\end{figure}

Furthermore, as illustrated in figure \ref{fig10}, the ratio of perpendicular to parallel energy and GCH dissipations at saturation decreases as $I$ increases.  When saturation is reached with a small value of $\chi$, perpendicular transfer and dissipation still strongly dominate the parallel ones. This is no longer the case when $I$ is large, as illustrated with the run $R_5$. Note that in this simulation, the initial value of $\chi$ was increased in order to shorten the time needed  to reach saturation. For such a large value of  $\chi$, the parallel dissipation rapidly overcomes the perpendicular one, as described in \citet{Meyrand21}.  It is to be noted that this situation is not specific of imbalanced turbulence but also occurs for balanced  simulations   at large $\chi$ (not shown), indicating that the breakdown of the asymptotics is not limited to imbalanced runs.
Moreover, it is conspicuous that the ratio of the perpendicular to parallel dissipation is always smaller for the GCH than for the energy, pointing out that the barrier acts dominantly on the GCH.

As the nonlinear parameter $\chi$ is increased,  the transition zone is less extended and displays a shallower slope (not shown),  except possibly when the imbalance is also increased, as in the simulations presented in \citet{Meyrand21}.

\subsection{Transition zone with negligible parallel dissipation}

\begin{figure}
\centerline{
\includegraphics[width=0.5\textwidth]{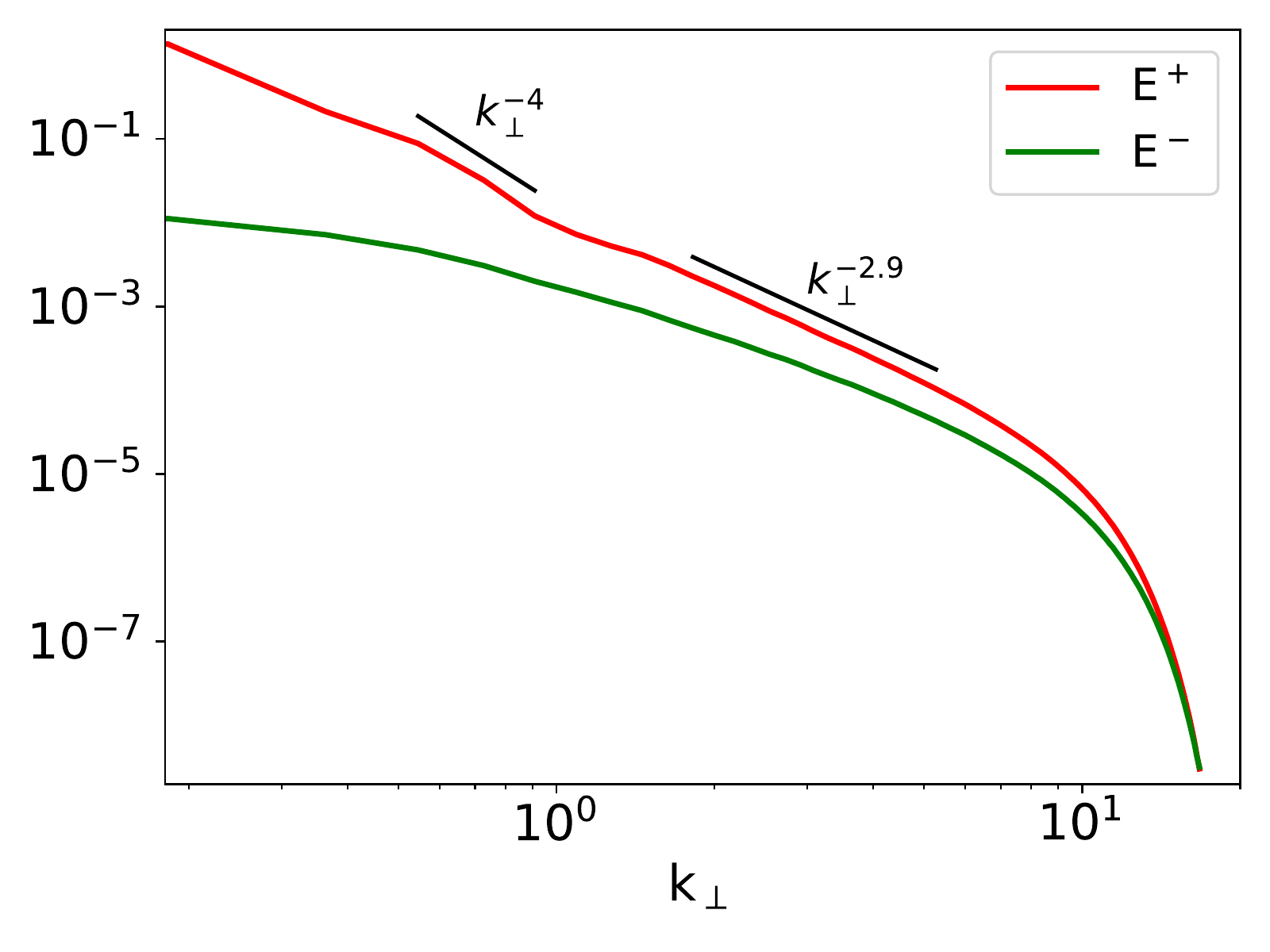}
\includegraphics[width=0.5\textwidth]{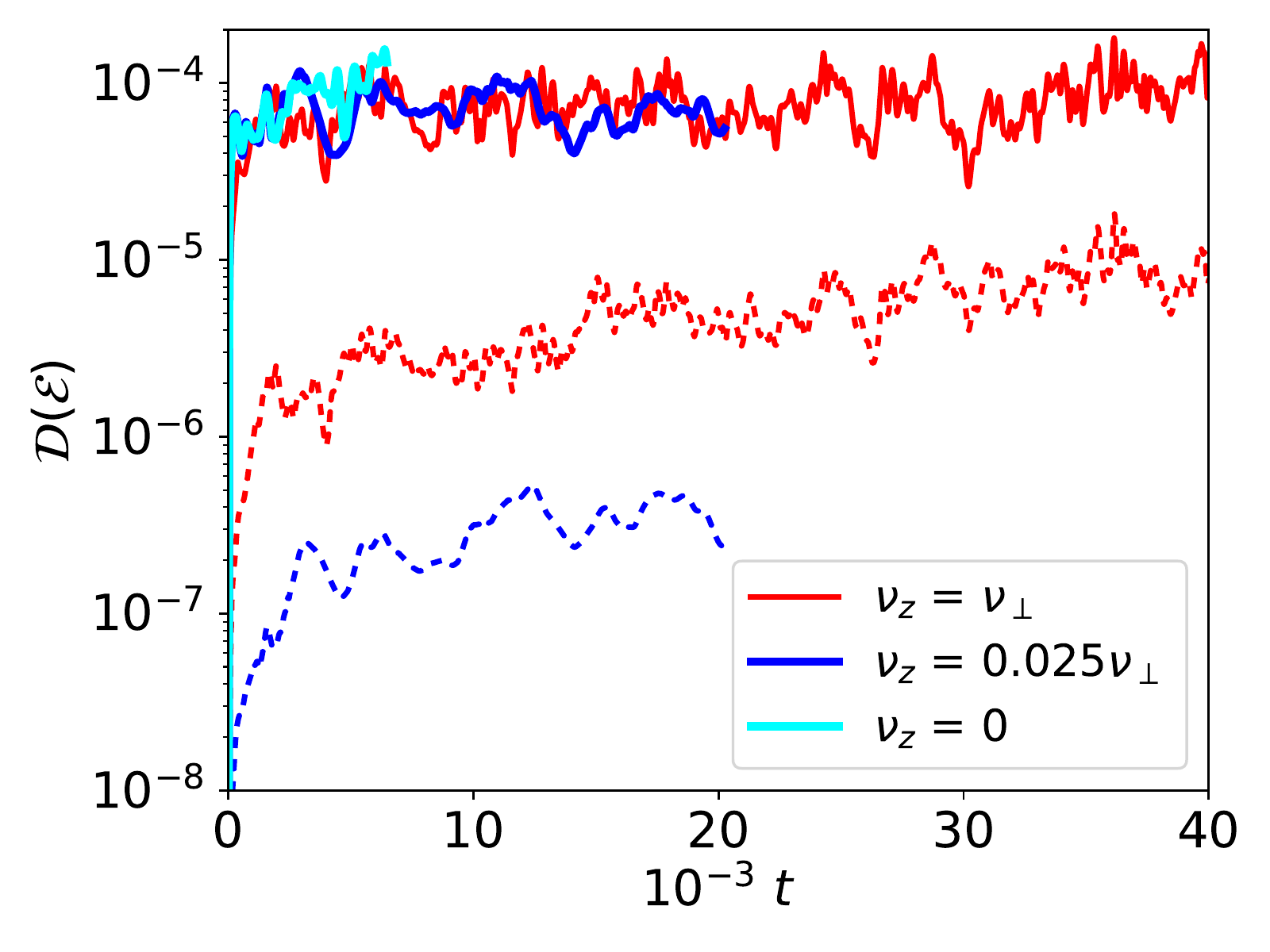}
}
\caption{Left panel: $E^\pm $ energy spectra, averaged over the time interval [15000, 20000], for run $R_6$ with a parallel viscosity smaller by a factor 40 than in run $R_2$. Right panel: perpendicular (solid lines) and parallel (dotted lines) energy dissipations (averaged over 3 Alfvén crossing times) for simulations performed with different parallel viscosities, all the other parameters being kept unchanged (red: run $R_2$; blue: run $R_6$; cyan: run $R_7$).}
\label{fig11}
\end{figure}

The transition zone not only exists in regimes where $\nu_z=\nu_\perp$ (as in all the runs presented in the previous sections), but persists when the parallel viscosity $\nu_z$ is decreased  while keeping $\nu_\perp$ unchanged. 
Figure \ref{fig11} (left) shows the transverse energy spectra $E^\pm(k_\perp)$ when $\nu_z$ is reduced by a factor $40$ relatively to its value in the run $R_2$. Parallel viscosity can in fact be taken arbitrary small and even zero  without significantly affecting the accuracy of the numerical simulation during a significant integration time. As seen on figure \ref{fig11} (right), when reducing $\nu_z$ the perpendicular dissipation remains unchanged, while the parallel one (which is always subdominant) decreases further and the contribution of the spectral range close to $k_\perp = 1$ still dominates that of the small transverse scales.
Regimes with small parallel dissipation are in fact relevant for the solar wind where the dynamics being quasi-transverse \citep{Sahraoui10}, the parallel transfer, and consequently the parallel dissipation  at the smallest scales, are much weaker than the transverse ones. 
The above results suggest that the existence of the transition zone is not a consequence of the parallel dissipation, at least for small values of the nonlinear parameter.  A possible origin of the transition in this case could be the interactions between co-propagating waves that take place at the sub-ion scales, as suggested by \citet{Voitenko16} and  \citet{Voitenko20}. This issue is addressed in the next section, using a semi-phenomenological model.

\section{Modeling of the transition zone}\label{transition-zone}

\subsection{A phenomenological model including co-propagating wave interactions}
\label{model-coprop}


\begin{figure}
\centering
\includegraphics[width=0.98\hsize]{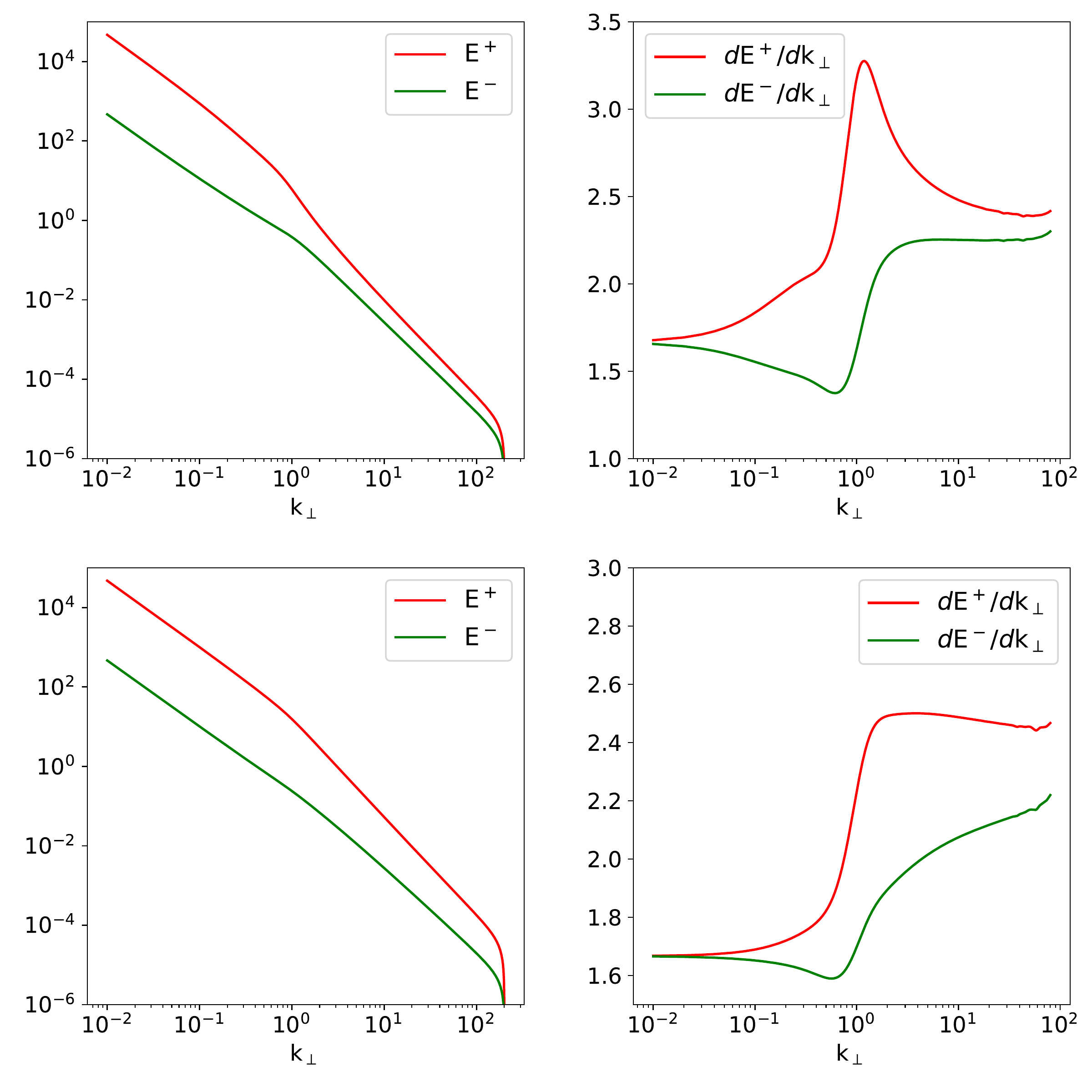}
\caption{Spectra $E^\pm((k_\perp)$ and local slopes predicted by the phenomenological model for $d=3$ (top) and $d=0$ (bottom), with  $C=15$, $\epsilon=1$, $\eta(k_\perp)=0.82 /v_{ph}^{1.1}(k_\perp)$.}
\label{fig12}
\end{figure}

We here concentrate on a diffusion model  in spectral space which was first derived in  \citet{PS19}, and then reproduced  in \citet{Milo20} using a heuristic approach based on the conservation of the energy and and of the GCH, together with a phenomenological estimate of the transfer times. This model does not capture the interactions between co-propagating waves, not only in the MHD range, where they do not exist, but also in the sub-ion range where co-propagative KAWs can actually interact. In \citet{PS19}, following \citet{Voitenko16}, we enriched the model by retaining interactions between triad wavenumbers which are comparable but not necessarily asymptotically close.
We introduce the nonlinear characteristic frequencies for the  mode ${\boldsymbol k}$ propagating in the $\pm$ direction,  $\gamma_{\boldsymbol k}^{\pm (\uparrow\uparrow)}= V_{\boldsymbol k}^{(\uparrow\uparrow)}  \sqrt{k_\perp^3 {E}^\pm(k_\perp)}$
and $\gamma_{\boldsymbol k}^{\pm (\uparrow\downarrow)}= V_{\boldsymbol k}^{(\uparrow\downarrow)} \sqrt{k_\perp^3 {E}^\mp(k_\perp)}$ 
(where $V_{\boldsymbol k}^{(\uparrow\uparrow)}$  and $V_{\boldsymbol k}^{(\uparrow\downarrow)} $ estimate the strength of the corresponding interactions),
associated with the interactions between co-propagating $(\uparrow\uparrow)$ or counter-propagating $(\uparrow\downarrow)$ waves, respectively.
Restricting the discussion to the case where both waves undergo strong nonlinear interactions,  we can define global inverse transfer times as
$(\tau_{tr, G}^\pm)^{-1} = \gamma_{\boldsymbol k}^{\pm (\uparrow\uparrow)} +\gamma_{\boldsymbol k}^{\pm(\uparrow\downarrow)}$. The ratio $\alpha(k_\perp)=V_{\boldsymbol k}^{(\uparrow\uparrow)} / V_{\boldsymbol k}^{(\uparrow\downarrow)}$ should be an increasing function of $k_\perp$, that is essentially zero in the MHD range and saturates to a finite value  in the dispersive range. 
Identifying the inverse transfer times $\gamma_{\boldsymbol k}^{\pm(\uparrow\downarrow)}$ associated with the contra-propagating waves 
with the nonlinear frequencies $\sqrt{k_\perp^3 v_{ph}^2 {E}^\mp(k_\perp)}$, the new inverse transfer times then rewrite
\begin{equation}
(\tau_{tr, G}^\pm)^{-1} \approx k^{3/2}_\perp v_{ph}
\left(({E}^\mp)^{1/2}+ \alpha(k_\perp)({E}^\pm)^{1/2}
\right).
\end{equation}
Defining $u^\pm = {E}^\pm/ k_\perp$, the equations governing the evolution of ${E}^\pm(k_\perp)$ is rewritten, in the stationary regime,
\begin{equation}
\frac{d}{d k_\perp} u^\pm(k_\perp)=-C\frac{\varepsilon\pm\eta v_{ph}}{k^5_\perp v_{ph}}\frac{1}
 {u^\mp(k_\perp)^{1/2}  + \alpha(k_\perp) u^\pm(k_\perp)^{1/2} },\label{eq:coprop-interaction} \\
\end{equation}
where the energy and GCH fluxes $\epsilon$ and $\eta$ are a priori functions of $k_\perp$ and $C$ is a numerical constant.
This equation is similar to Eq.(7.34) of \citet{PS19} when $\nu=0$, except that the weighting factor of the co-propagating wave interactions is here taken to be $\alpha(k)$ instead of its square.

 The function $\alpha$, which  is taken as $\alpha(k_\perp)=3((1+ck_\perp^2)^{1/2}-1)/((1+ck_\perp^2)^{1/2}+1)$   in \citet{Voitenko16},  with $c$  denoting a slowly varying function of $k_\perp$, can also be written
\begin{equation}
\alpha(k_\perp)=d\left ( \frac{v_{ph}(k_\perp)-s}{v_{ph}(k_\perp)+s}\right ),
\end{equation}
where $s = \sqrt{2/\beta_e}$ is the Alfv\'en velocity (in sound speed units) and  $d$ a free constant parameter.
An important remark, related to the existence of the helicity barrier, is that Eq. (\ref{eq:coprop-interaction}) does not have a satisfactory solution when $\eta\ne 0$. The linear growth of the phase velocity $v_{ph}$ at small scales implies that for $k_\perp$ large enough, depending on the sign of $\eta$, one of the spectra ${E}^+$ or ${E}^-$ increases with $k_\perp$. 

\subsection{Numerical integration of the  phenomenological model}
\label{numeric-model}
As discussed in \citet{PS19}, Eq. (\ref{eq:coprop-interaction}) can be considered either as an initial value problem when the energy and generalized cross-helicity fluxes $\varepsilon$ and $\eta$  are given and the spectra ${E}^\pm$  prescribed at a wavenumber $k_\perp=k_0$ or, alternatively, as
a nonlinear eigenvalue problem for $\varepsilon$ and $\eta$, when the spectra are specified at  $k_0$ and prescribed to decrease to zero at infinity. Numerical integration of these equations is best performed as an initial value problem where $k_0$ is taken close to the largest wavenumber of the spectral domain. Indeed, the eigenvalue problem (where ideally $k_0$ is chosen at the lower end of the spectral domain) requires a shooting method and turns out to be extremely difficult to solve numerically, due to its instability. Indeed, under the effect of any small perturbation, the small-scale spectrum takes the form of an absolute equilibrium or goes to zero at a finite value of $k_\perp$, depending on the boundary conditions. 

In the integrations presented below, we choose $\beta=2$, $\tau=1$, $k_0=200$ and small enough initial conditions such as to obtain a solution where the spectra tend to zero for a value of $k_\perp$ slightly in excess of $k_0$. As suggested by the numerical simulations  of the gyrofluid equations, in the conditions where $\chi$ is not too large, $\varepsilon$ can be taken constant. 
We used $\varepsilon=1$ (close to the almost constant energy flux observed in the gyrofluid simulations) and $C=15$. Differently, the gyrofluid simulations show that the flux of GCH undergoes a significant decrease at the barrier \footnote{As discussed in \citet{PS19} and \citet{Milo20}, a constant GCH flux would lead to unrealistically large amplitudes and large imbalance at large scale, a consequence of the fact that transfer in the z-direction is not retained by the diffusion model.} To comply with this observation, we were led to prescribe that $\eta$ decreases faster than $1/v_{ph}(k_\perp)$. 
For simplicity, we here chose to take $\eta(k_\perp)=\eta_0/v_{ph}^{\gamma}(k_\perp)$. The numerical factor $\eta_0=0.82$ ensures that the imbalance at large scales tends to a value close to $100$, as in most of the numerical simulations presented in Section \ref{GCH-barrier}, while the exponent $\gamma=1.1$ leads to a small-scale imbalance which is also compatible with the simulations. A larger value of $\gamma$ would produce a smaller imbalance at small scales. Note that decaying transfer rates also arise in nonlinear diffusion models of turbulence retaining Landau damping \citep{Howes11}.

In figure \ref{fig12}, we display the ${E}^\pm$ spectra (left) and their local spectral exponents (right) for the cases $d=3$ (top) and $d=0$ (bottom), the other parameters being taken as mentioned above.  In the absence of co-propagating wave interactions ($d=0$), the ${E}^+$ spectrum exponent undergoes a transition, at $k_\perp$ close to unity, between the Kolmogorov exponent $-5/3$ and a value slightly steeper than the theoretical value $-7/3$,  as expected in the presence of imbalance \citep{PS19}. When $d=3$ (a value compatible with kinetic theory, used by \citet{Voitenko16}), a clear overshoot of the spectral index is observed, associated with the existence of a transition zone which starts for $k_\perp$ slightly smaller than unity, the steeper slope being equal to -3.28 at $k_\perp=1.18$. Near this wavenumber, the ${E}^-$ spectrum displays a weak flattening.

Several remarks can be made concerning the influence of the various parameters of the model, which, at this stage, remains only qualitative. First, the slope of the transition zone is steeper (shallower) as $d$ in increased (respectively decreased). This further indicates the role of the co-propagating wave interactions in the generation of this transition zone. 

At larger imbalance (e.g. when taking $\eta_0=0.92$, which leads to a large-scale imbalance close to $I=1000$), the  transition zone is slightly steeper, with a slope reaching $3.52$ at  $k_\perp=1.11$.

In the absence of co-propagating wave interactions, and when the variation of $\eta$ is such that $\eta v_{ph}(k_\perp)$ is constant, the imbalance remains independent of $k_\perp$.  This is no longer the case when $\eta$ decreases faster than $1/v_{ph}(k_\perp)$. As $\gamma$ is increased from the value $1.1$ chosen in the figure, the imbalance at small scale is strongly reduced but the transition zone takes a different form. While the slope of the ${E}^+$ spectrum is not as steep as for smaller values of $\gamma$, the ${E}^-$ spectrum becomes shallower.  The case where $\eta$ decreases faster can 
be associated with a stronger transfer and/or dissipation in the $z$-direction and thus a larger value of $\chi$. This effect is seen in some simulations but a systematic study has not been performed.

\section{Conclusion}  \label{conclusion}
Using a two-field gyrofluid model retaining  AWs only, we have observed that, for $\beta_e$ of order unity or larger and a weak or moderate energy imbalance,  a transition zone can exist in the absence of a significant parallel energy dissipation, in contrast with the mechanism based on the dissipation induced by the ion-cyclotron resonance,  recently proposed by \citet{Squire21}. This transition zone is found to be steeper at larger imbalance, in line with Parker Solar Probe observations \citep{Huang_2021}.
In these simulations, an helicity barrier also forms, associated with a decay of the perpendicular CGH flux at the ion scales, while the corresponding energy flux is almost scale independent.
Such simulations are consistent with the possible effect of the interactions between co-propagating AWs. The efficiency of this mechanism is supported by the analysis of a phenomenological model based on diffusion equations in the spectral space, in the spirit of the Leith's model \citep{Leith67} for hydrodynamic turbulence. 
It turns out that, in the presence of co-propagative waves, and only in this case, the energy barrier, modeled by a decay of the GCH flux with the transverse wavenumber, leads to a spectral transition zone. As a consequence of weak or negligible parallel dissipation, the injected energy accumulates and a relatively long time is needed for turbulence to reach a steady state.  As a consequence, in the context of a more realistic model, one may suspect that before ion-cyclotron wave can be excited, other dissipative effects can come into play, and among them, Landau damping or stochastic heating. Kinetic simulations for values of the beta parameters of order unity or larger are necessary to address these issues.

\section*{Acknowledgments}
This work was granted access to the HPC resources of CINES/IDRIS under the allocation A0090407042. Part of the computations have also been done on the “Mesocentre SIGAMM” machine, hosted by Observatoire de la Côte d’Azur.

\section*{Declaration of Interests}
The authors report no conflict of interest.

\bibliographystyle{jpp}
\bibliography{biblio}

\end{document}